\documentclass[twocolumn]{aastex63}
\usepackage{amsmath}
\usepackage{soul}

\newcommand{\lya}{{Ly$\alpha$}}
\newcommand{\oii}{[O\,{\sc ii]} $\lambda$3727}
\newcommand{\hb}{H$\beta$}
\newcommand{\oiii}{[O\,{\sc iii]} $\lambda$5007}
\newcommand{\ha}{H$\alpha$}
\newcommand{\hi}{H\,{\sc i}}
\newcommand{\hii}{H\,{\sc ii}}
\newcommand{\hf}{$x_{\rm HI}$}

\newcommand{\pj}{J17}
\newcommand{\pn}{N20}
\newcommand{\pz}{Z21}

\received{2021 Oct 20}
\revised{2021 Dec 08}
\accepted{2021 Dec 11}

\shorttitle{\lya\ Emitters and LF at $z \approx 6.6$}
\shortauthors{Ning et al.}

\begin{document}

\title{\large The Magellan M2FS Spectroscopic Survey of High-$z$ Galaxies: \lya\ Emitters at $z\approx6.6$ and the Evolution of \lya\ Luminosity Function over $z\approx5.7-6.6$}

\author[0000-0001-9442-1217]{Yuanhang Ning}
\altaffiliation{ningyhphy@pku.edu.cn}
\affiliation{Kavli Institute for Astronomy and Astrophysics, Peking University, Beijing 100871, China}
\affiliation{Department of Astronomy, School of Physics, Peking University, Beijing 100871, China}

\author[0000-0003-4176-6486]{Linhua Jiang}
\altaffiliation{jiangKIAA@pku.edu.cn}
\affiliation{Kavli Institute for Astronomy and Astrophysics, Peking University, Beijing 100871, China}
\affiliation{Department of Astronomy, School of Physics, Peking University, Beijing 100871, China}

\author[0000-0002-9634-2923]{Zhen-Ya Zheng}
\affiliation{CAS Key Laboratory for Research in Galaxies and Cosmology, Shanghai Astronomical Observatory, Shanghai 200030, China}

\author[0000-0001-5364-8941]{Jin Wu}
\affiliation{Kavli Institute for Astronomy and Astrophysics, Peking University, Beijing 100871, China}
\affiliation{Department of Astronomy, School of Physics, Peking University, Beijing 100871, China}

\begin{abstract}
We present a sample of \lya\ emitters (LAEs) at $z\approx6.6$ from our spectroscopic survey of high-redshift galaxies using the multi-object spectrograph M2FS on the Magellan Clay telescope. The sample consists of 36 LAEs selected by the narrow-band (NB921) technique over nearly 2 deg$^2$ in the sky. These galaxies generally have high \lya\ luminosities spanning a range of ${\sim}3\times10^{42}{-}7\times10^{43}$ erg~s$^{-1}$, and include some of the most \lya-luminous galaxies known at this redshift. They show a positive correlation between the \lya\ line width and \lya\ luminosity, similar to the relation previously found in $z\approx5.7$ LAEs. Based on the spectroscopic sample, we calculate a sophisticated sample completeness correction and derive the \lya\ luminosity function (LF) at $z\approx6.6$. We detect a density bump at the bright end of the \lya\ LF that is significantly above the best-fit Schechter function, suggesting that very luminous galaxies tend to reside in overdense regions that have formed large ionized bubbles around them. By comparing with the $z\approx5.7$ \lya\ LF, we confirm that there a rapid LF evolution  at the faint end, but a lack of evolution at the bright end. The fraction of the neutral hydrogen in the intergalactic medium at $z\approx6.6$ estimated from such a rapid evolution is about $\sim0.3\pm0.1$, supporting a rapid and rather late process of cosmic reionization.
\end{abstract}


\section{Introduction}

During the epoch of cosmic reionization (EoR), neutral hydrogen (\hi) in the intergalactic medium (IGM) was ionized by ultraviolet (UV) radiation from early astrophysical objects. The fraction of \hi\ (\hf) in the IGM is a key quantity to depict the history of EoR. The Gunn-Peterson effect in high-redshift quasar spectra has been used to investigate highly ionized IGM with \hf\ $\lesssim10^{-4}$, and has suggested that cosmic reionization ended at $z\lesssim6$ \citep{fan06}. Meanwhile, high-redshift star-forming galaxies can also play an important role on estimating \hf\ with a wide range of ${\sim}0.1{-}1$ \citep{mr04, kashikawa06, kashikawa11, ouchi10}. The \lya\ emission line is a good tracer to search for star-forming galaxies at high redshift \citep{pp67}. These galaxies, known as \lya\ emitters (LAEs), are routinely found to study the EoR and the properties of high-redshift galaxies \citep[e.g.,][]{kashikawa06, kashikawa11, hu10, finkelstein12, jiang13a, jiang16a, jiang20a, pentericci14, zheng16, ota17}.

LAE candidates are usually selected in ground-based narrowband imaging surveys, and follow-up spectroscopy is sometimes taken to confirm them. This narrowband technique has been widely used to search for high-redshift LAEs at $z\approx 5.7$, 6.6, and 7.0 \citep[e.g.,][]{hu02, kodaira03, rhoads04, taniguchi05, iye06, kashikawa06, kashikawa11, hu10, rhoads12, jiang17, zheng17, shibuya18a}. Some LAEs (or candidates) at $z>7$ have also been reported using this technique \citep[e.g.,][]{tilvi10, shibuya12}. These LAEs constitute unique galaxy samples to study the early universe. 

Previous studies based on different LAE samples show a rapid evolution of the \lya\ luminosity function (LF) from $z\approx5.7$ to 6.6 \citep[e.g.,][]{kashikawa06, kashikawa11, ouchi08, hu10, henry12, konno18}. However, significant discrepancies exist in the measurements of the \lya\ LFs, not only between the results from photometrically selected samples and spectroscopically confirmed samples \citep[e.g.,][]{matthee15, bagley17}, but also among different spectroscopic samples \citep{kashikawa06, kashikawa11, hu10, ouchi10}. The discrepancies on LF normalizations can be as large as a factor of 2 to 3. The reason for these discrepancies is unclear, but may include target contamination, sample incompleteness, and cosmic variance. We thus need to build a spectroscopically confirmed LAE sample with a high completeness over a large sky area.

In this paper we present a sample of spectroscopically confirmed LAEs at $z\approx6.6$ from a survey program. The survey was designed to build a large and homogeneous sample of high-redshift galaxies, including LAEs at $z\approx5.7$ and 6.6, and Lyman-break galaxies (LBGs) at $5.5<z<6.8$. We carried out spectroscopic observations using the fiber-fed, multi-object spectrograph Michigan/Magellan Fiber System (M2FS; \citealt{mateo12}) on the 6.5 m Magellan Clay telescope. Our targets come from five well-studied fields with a total sky area about 2 deg$^2$, including the Subaru {\it XMM-Newton} Deep Survey (SXDS), A370, the Extended {\it Chandra} Deep Field-South (ECDFS), COSMOS, and SSA22. \citet[][hereafter \pj]{jiang17} provided an overview about the program. From the survey, we have built a large sample of 260 LAEs at $z\approx5.7$ \citep[][hereafter \pn]{ning20}. Based on the sample, Zheng et al. (2021, hereafter \pz) calculated a \lya\ LF at $z\approx5.7$. Here we provide a sample of 36 LAEs at $z\approx6.6$ and construct a \lya\ LF at this redshift.

The paper has a layout as follows. In Section 2, we briefly review the M2FS survey, including spectroscopic observations, data reduction, and target selection. In Section 3, we present the spectroscopically confirmed LAEs at $z\approx6.6$ and their spectra. We measure the sample completeness and present the \lya\ LF at $z\approx6.6$ in Section 4.  In Section 5, we discuss the evolution of the \lya\ LF at high redshift and its implication. We summarize our paper in Section 6. Throughout the paper, we use a standard flat cosmology with $H_0=\rm{70\ km\ s^{-1}\ Mpc^{-1}}$, $\Omega_m=0.3$ and $\Omega_{\Lambda}=0.7$. All magnitudes refer to the AB system.

\section{Survey Outline and Target Selection}

\subsection{Survey Outline}

In this survey, we used Magellan M2FS to carry out spectroscopic observations of galaxies. The scientific goal is to build a large and homogeneous sample of high-redshift LAEs and LBGs. Based on this sample, we can study properties of these galaxies, \lya\ LF and its evolution at high redshift, high-redshift protoclusters, cosmic reionization, etc. The galaxy candidates were selected from five fields, including SXDS, A370, ECDFS, COSMOS, and SSA22. These fields have deep optical images in a series of broad $[BVR(r')I(i')z']$ and narrow bands (e.g., NB816 and NB921) from Subaru Suprime-Cam. The fields are summarized in Table \ref{fieldsinfo}. Since the depths of the SXDS images slightly vary across the five Suprime-Cam pointings, the five pointings are shown as SXDS1-5, respectively. We treat them as five different fields in this work, although they have marginal overlapping regions. Columns 4 and 5 list the magnitude limits of the $z'$ and NB921-band images used for our candidate selection of LAEs at $z\approx6.6$. The average depth ($5\sigma$ detections in a $2\arcsec$-diameter aperture) is ${\sim}26.2$ mag in $z'$, and ${\sim}25.5$ mag in NB921.

\begin{deluxetable}{ccccc}
\tablecaption{Survey Fields
\label{fieldsinfo}}
\tablewidth{0pt}
\tablehead{
\colhead{Field} & \colhead{Coordinates} & \colhead{Area} & \colhead{$z'$} & \colhead{NB921}\\
\colhead{} & \colhead{(J2000.0)} & \colhead{(deg$^2$)} & \colhead{(mag)} & \colhead{(mag)}
}
\colnumbers
\startdata
	SXDS1	&	02:18:18.2 --05:00:09.96	&	0.1743	&	26.0	&	25.2	\\
	SXDS2	&	02:17:47.8 --04:35:26.63	&	0.1795	&	26.1	&	25.6	\\
	SXDS3	&	02:17:46.0 --05:26:17.88	&	0.1758	&	25.8	&	25.6	\\
	SXDS4	&	02:19:43.5 --05:01:39.25	&	0.1777	&	25.9	&	25.5	\\
	SXDS5	&	02:16:16.6 --05:00:45.04	&	0.1775	&	26.0	&	25.6	\\
	A370a	&	02:39:49.4 --01:35:12.16	&	0.1692	&	26.2	&	25.8	\\
	ECDFS	&	03:31:59.8 --27:49:17.07	&	0.1386	&	27.1	&	25.8	\\
	COSMOS	&	10:00:29 +02:12:21		&	0.3909	&	26.2	&	25.4	\\
	SSA22a	&	22:17:26.5 +00:13:40.89	&	0.1664	&	26.3	&	25.3	\\
\enddata
\tablecomments{Column 1 lists the field names. Column 2 shows the coordinates of the M2FS pointing centers. For the COSMOS field (five pointings, COSMOS1-5), we provide its center of the Suprime-Cam observations. Column 3 gives the area of the fields covered by M2FS pointings. COSMOS is given by the coverage area of the three COSMOS (2, 4, and 5) pointings. Columns 4 and 5 indicate the magnitude limits ($5\sigma$ detections in a $2\arcsec$-diameter aperture).}
\end{deluxetable}

M2FS has a field-of-view of about half a degree in diameter and high efficiency to detect relatively bright, high-redshift galaxies. The fibers have an angular diameter of $1.2\arcsec$, significantly larger than the sizes of galaxies at $z\ge6$. We used a pair of standard 600-line red-blazed reflection gratings. The resolving power is about 2000, and the wavelength coverage is roughly from 7600 to 9600 \AA. In Table \ref{fieldsinfo}, Column 2 shows the coordinates of the M2FS pointing centers. Note that COSMOS lists its central coordinate of the Suprime-Cam field because it has five M2FS pointings. The LAE candidates at $z\approx5.7$ and 6.6 in the M2FS pointings were all observed due to their higher priorities in our M2FS program. In addition, each pointing also covered $z\gtrsim6$ LBG candidates, a variety of ancillary targets, several bright reference stars, and a few tens (typically around 50) of sky fibers.

We have completed the M2FS observations and reduced the spectroscopic data. The effective integration time per pointing was about 5 hr on average. We used our own customized pipeline for data reduction. The pipeline can produce both one-dimensional (1D) and two-dimensional (2D) spectra. It performs bias (overscan) correction, dark subtraction, flat-fielding, cosmic ray identification, production of ``calibrated" 2D images, 1D-spectra extraction by tracing fiber positions in twilight images, wavelength calibration using the 1D lamp spectra (or OH-skyline forests in the science spectra), and sky subtraction by averaging spectra of sky fibers. The pipeline can also produce the 2D spectra of individual exposures for visual inspection and comparison. 

\subsection{Target Selection}

\begin{figure}[t]
\epsscale{1.15}
\plotone{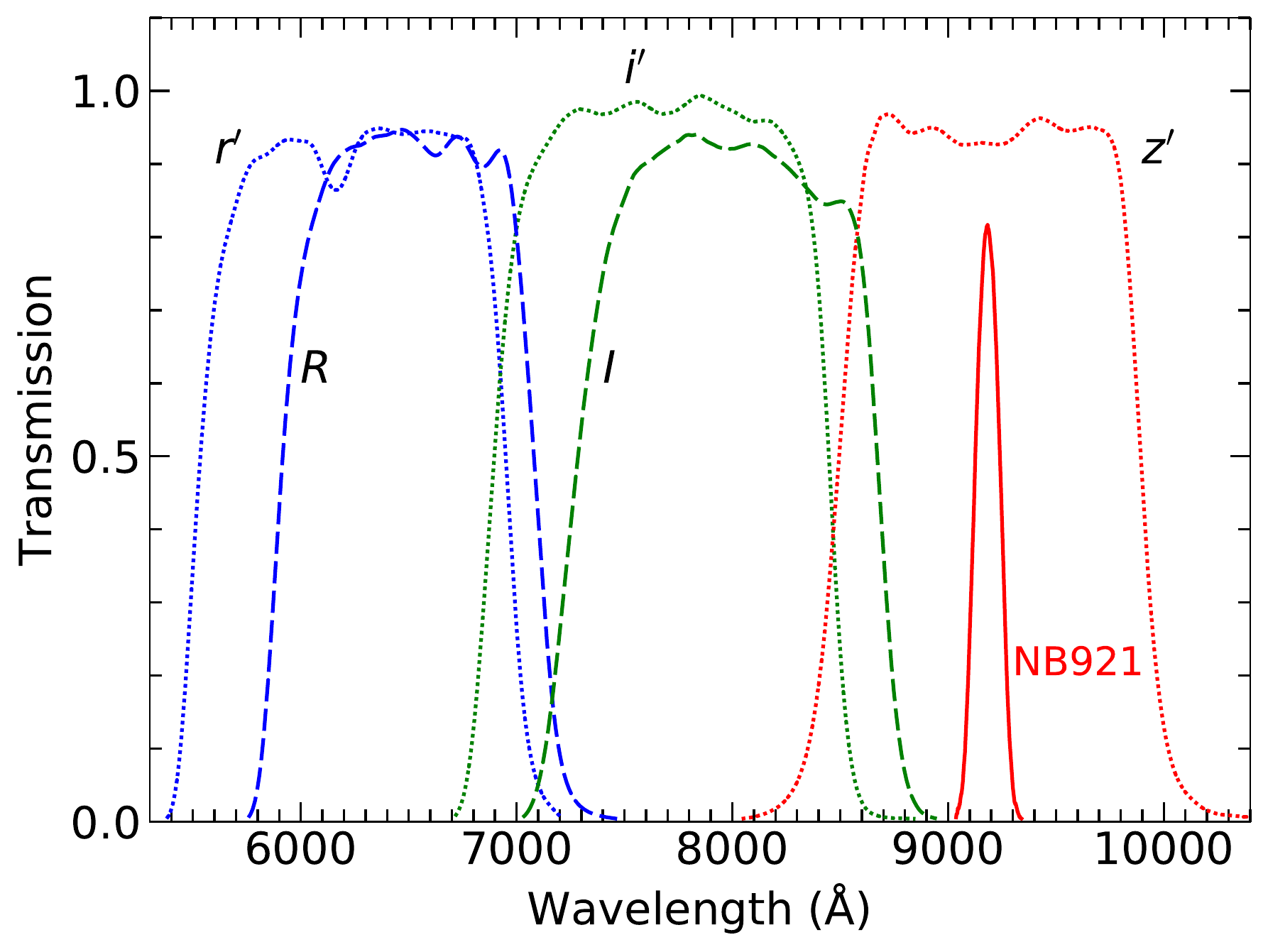}
\caption{Transmission curves of the Suprime-Cam filters that were used for 
our target selection. The NB921 band corresponds to the detection of LAEs 
at $z\approx6.6$.
\label{filters}}
\end{figure}

LAE candidates are usually selected using the narrowband (or \lya) technique. In Figure \ref{filters}, we show the transmission curves of the filters used for our selection of LAEs at $z\approx6.6$. Different fields have slightly different combinations of the broadband filters, such as $r'i'z'$, $Ri'z'$, and $RIz'$. We mainly used the $z'-{\rm NB921}$ color to select the candidates of LAEs at $z\approx6.6$. For all $>5\sigma$ detections in the NB921 band, we applied the following color cut, 
\begin{eqnarray}
   z'-{\rm NB921}>0.8.
\end{eqnarray}
The criterion is similar to those used in the literature \citep[e.g.,][]{hu10, ouchi10, kashikawa11}. It roughly corresponds to a line equivalent width (EW) of ${\gtrsim}20$ \AA\ in the rest frame. Objects undetected in $z'$ band also satisfy the color selection, because the $z'$-band images are much deeper than the NB921-band images (see Table \ref{fieldsinfo}). The color-magnitude diagram for target selection of the $z\approx6.6$ candidates in SXDS is illustrated by the middle panel in Figure 7 of \pj.

Another two criteria were also applied to eliminate lower-redshift contaminants. We required that candidates should not be detected ($<2\sigma$) in $B$ or $V$ band, assuming that no flux can be detected in wavelength range bluer than the Lyman limit. We also applied color selection of $i' (I)-z' > 1.2 (1.0)$ and $r' (R)-z' > 2.5$ for objects detected (${>}3\sigma$) in $z'$. Note that our images typically reach $\sim28$ mag in the $BV$ bands and $\sim27.5$ mag in the $Rr'Ii'$ bands ($3\sigma$ in a 2\arcsec\ diameter aperture).
The above two criteria do not remove real objects at $z\approx6.6$. We also visually inspected all candidates. We have removed spurious detections such as satellite trails and residuals of bright star spikes. We have also removed objects whose photometric measurements are largely influenced by nearby bright stars.

For each field, the selection of M2FS pointing centers was restricted by the number and spatial distribution of bright stars therein. Although not all $z\approx6.6$ LAE candidates in the field were observed (see Figure 1--5 of \pj), those inside the M2FS pointings were observed as we introduce above. Therefore, the valid region for each field is the region covered by the M2FS pointings. In Table \ref{fieldsinfo}, Column 3 gives the covered areas of the fields by the M2FS pointings. The field-of-view of M2FS is $29\arcmin.2$ in diameter. The area values are then used in our \lya\ LF calculation later.

\renewcommand{\arraystretch}{1}
\floattable
\begin{deluxetable}{ccccccccc}
\tablecaption{36 LAEs in the $z\approx6.6$ sample
\label{sample}}
\tablehead{
   \colhead{No.} & \colhead{R.A.} & \colhead{Decl.} & \colhead{Redshift} & \colhead{$z'$} & \colhead{NB921} & \colhead{$F_{\rm cont}$} & \colhead{${\rm log}_{10} L$(\lya)} & \colhead{$V_a$}\\ 
   \colhead{} & \colhead{(J2000.0)} & \colhead{(J2000.0)} & \colhead{} & \colhead{(mag)} & \colhead{(mag)} & \colhead{($10^{-19}\ \rm erg\ s^{-1}cm^{-1}\AA^{-1}$)} & \colhead{($\rm erg\ s^{-1}$)} & \colhead{($10^6$ cMpc$^3$)}
   }
\colnumbers
\startdata
\vspace{0.05cm}
 01 &  02:39:08.54 &  $-$01:31:26.2 &        6.492 &  27.06 $\pm$ 0.47 &  25.11 $\pm$ 0.11 &          $<${0.95} &  ${43.22}_{-0.06}^{+0.05}$ &  ${1.507}_{-0.103}^{+0.067}$ \\\vspace{0.05cm}
 02 &  02:19:30.84 &  $-$05:07:19.1 &        6.498 &           $>$26.9 &  25.25 $\pm$ 0.16 &          $<${0.42} &  ${43.02}_{-0.13}^{+0.11}$ &  ${1.039}_{-0.435}^{+0.302}$ \\\vspace{0.05cm}
 03 &  02:19:26.97 &  $-$05:07:21.4 &        6.505 &  26.19 $\pm$ 0.28 &  25.22 $\pm$ 0.16 &  ${1.08}\pm{0.46}$ &  ${42.66}_{-0.41}^{+0.28}$ &  ${0.069}_{-0.067}^{+0.703}$ \\\vspace{0.05cm}
 04 &  03:31:48.55 &  $-$27:53:52.4 &        6.510 &  26.21 $\pm$ 0.09 &  24.82 $\pm$ 0.08 &  ${0.80}\pm{0.15}$ &  ${42.91}_{-0.11}^{+0.08}$ &  ${0.671}_{-0.340}^{+0.270}$ \\\vspace{0.05cm}
 05 &  02:18:27.02 &  $-$04:35:08.0 &        6.514 &  25.52 $\pm$ 0.12 &  24.24 $\pm$ 0.06 &  ${1.73}\pm{0.33}$ &  ${43.03}_{-0.13}^{+0.10}$ &  ${1.071}_{-0.434}^{+0.270}$ \\\vspace{0.05cm}
 06 &  02:16:54.38 &  $-$05:00:04.4 &        6.514 &           $>$27.0 &  24.78 $\pm$ 0.10 &          $<${0.15} &  ${43.13}_{-0.06}^{+0.03}$ &  ${1.341}_{-0.152}^{+0.063}$ \\\vspace{0.05cm}
 07 &  02:18:43.65 &  $-$05:09:15.7 &        6.515 &  26.01 $\pm$ 0.21 &  24.66 $\pm$ 0.13 &  ${1.06}\pm{0.37}$ &  ${42.94}_{-0.21}^{+0.15}$ &  ${0.772}_{-0.601}^{+0.472}$ \\\vspace{0.05cm}
 08 &  02:18:23.54 &  $-$04:35:24.1 &        6.519 &  26.10 $\pm$ 0.21 &  25.21 $\pm$ 0.14 &  ${1.23}\pm{0.33}$ &  ${42.48}_{-0.47}^{+0.23}$ &  ${0.009}_{-0.009}^{+0.127}$ \\\vspace{0.05cm}
 09 &  02:17:14.01 &  $-$05:36:48.8 &        6.530 &  24.66 $\pm$ 0.09 &  23.64 $\pm$ 0.04 &  ${4.48}\pm{0.51}$ &  ${43.03}_{-0.13}^{+0.11}$ &  ${1.071}_{-0.434}^{+0.292}$ \\\vspace{0.05cm}
 10 &  02:18:33.96 &  $-$05:18:37.3 &        6.532 &  26.17 $\pm$ 0.28 &  25.05 $\pm$ 0.12 &  ${1.08}\pm{0.40}$ &  ${42.67}_{-0.28}^{+0.15}$ &  ${0.079}_{-0.075}^{+0.308}$ \\\vspace{0.05cm}
 11 &  02:39:39.46 &  $-$01:34:32.7 &        6.540 &  26.91 $\pm$ 0.42 &  24.69 $\pm$ 0.07 &          $<${0.59} &  ${43.05}_{-0.06}^{+0.04}$ &  ${1.130}_{-0.189}^{+0.113}$ \\\vspace{0.05cm}
 12 &  02:40:01.82 &  $-$01:41:00.2 &        6.544 &           $>$27.2 &  24.69 $\pm$ 0.08 &          $<${0.09} &  ${43.06}_{-0.05}^{+0.02}$ &  ${1.160}_{-0.154}^{+0.055}$ \\\vspace{0.05cm}
 13 &  10:01:24.80 &  $+$02:31:45.4 &        6.544 &  25.78 $\pm$ 0.16 &  23.76 $\pm$ 0.04 &  ${0.80}\pm{0.32}$ &  ${43.36}_{-0.03}^{+0.03}$ &                          --- \\\vspace{0.05cm}
 14 &  02:20:23.83 &  $-$05:06:48.9 &        6.551 &           $>$26.9 &  25.19 $\pm$ 0.15 &          $<${0.51} &  ${42.77}_{-0.09}^{+0.08}$ &  ${0.255}_{-0.164}^{+0.223}$ \\\vspace{0.05cm}
 15 &  02:16:11.39 &  $-$04:56:33.4 &        6.551 &           $>$27.0 &  24.96 $\pm$ 0.12 &          $<${0.35} &  ${42.91}_{-0.06}^{+0.05}$ &  ${0.671}_{-0.193}^{+0.169}$ \\\vspace{0.05cm}
 16 &  09:58:55.04 &  $+$01:53:41.8 &        6.552 &           $>$27.2 &  24.83 $\pm$ 0.14 &          $<${0.21} &  ${42.98}_{-0.06}^{+0.05}$ &  ${0.909}_{-0.204}^{+0.162}$ \\\vspace{0.05cm}
 17 &  02:19:01.44 &  $-$04:58:59.0 &        6.554 &  26.39 $\pm$ 0.32 &  24.35 $\pm$ 0.07 &          $<${0.72} &  ${43.16}_{-0.05}^{+0.04}$ &  ${1.404}_{-0.111}^{+0.072}$ \\\vspace{0.05cm}
 18 &  02:18:27.02 &  $-$05:07:27.1 &        6.557 &  26.67 $\pm$ 0.36 &  25.10 $\pm$ 0.18 &          $<${0.64} &  ${42.79}_{-0.12}^{+0.12}$ &  ${0.304}_{-0.225}^{+0.367}$ \\\vspace{0.05cm}
 19 &  02:19:31.78 &  $-$05:06:15.5 &        6.558 &  25.97 $\pm$ 0.21 &  24.27 $\pm$ 0.07 &  ${0.95}\pm{0.35}$ &  ${43.15}_{-0.06}^{+0.04}$ &  ${1.384}_{-0.140}^{+0.076}$ \\\vspace{0.05cm}
 20 &  02:17:51.43 &  $-$04:24:49.9 &        6.559 &           $>$27.1 &  25.50 $\pm$ 0.18 &          $<${0.45} &  ${42.65}_{-0.11}^{+0.09}$ &  ${0.059}_{-0.043}^{+0.130}$ \\\vspace{0.05cm}
 21 &  02:17:35.78 &  $-$04:25:24.7 &        6.564 &           $>$27.1 &  25.15 $\pm$ 0.13 &          $<${0.35} &  ${42.85}_{-0.06}^{+0.06}$ &  ${0.478}_{-0.173}^{+0.193}$ \\\vspace{0.05cm}
 22 &  02:20:26.83 &  $-$05:05:42.6 &        6.565 &           $>$26.9 &  24.78 $\pm$ 0.10 &          $<${0.32} &  ${43.02}_{-0.05}^{+0.04}$ &  ${1.039}_{-0.163}^{+0.122}$ \\\vspace{0.05cm}
 23 &  22:17:59.68 &  $+$00:22:31.2 &        6.569 &           $>$27.3 &  25.02 $\pm$ 0.15 &          $<${0.21} &  ${42.95}_{-0.06}^{+0.06}$ &  ${0.806}_{-0.202}^{+0.201}$ \\\vspace{0.05cm}
 24 &  09:59:35.08 &  $+$02:25:05.2 &        6.573 &  26.67 $\pm$ 0.35 &  24.99 $\pm$ 0.14 &          $<${0.62} &  ${42.94}_{-0.09}^{+0.07}$ &  ${0.772}_{-0.294}^{+0.235}$ \\\vspace{0.05cm}
 25 &  10:00:18.35 &  $+$02:00:08.2 &        6.574 &  26.38 $\pm$ 0.23 &  24.81 $\pm$ 0.14 &  ${0.68}\pm{0.27}$ &  ${42.99}_{-0.08}^{+0.07}$ &  ${0.941}_{-0.270}^{+0.219}$ \\\vspace{0.05cm}
 26 &  02:18:06.23 &  $-$04:45:10.8 &        6.577 &  26.71 $\pm$ 0.36 &  24.28 $\pm$ 0.06 &          $<${0.62} &  ${43.29}_{-0.03}^{+0.02}$ &  ${1.596}_{-0.033}^{+0.023}$ \\\vspace{0.05cm}
 27 &  02:19:33.14 &  $-$05:08:20.8 &        6.592 &  26.57 $\pm$ 0.38 &  24.96 $\pm$ 0.12 &          $<${0.76} &  ${43.12}_{-0.07}^{+0.05}$ &  ${1.317}_{-0.187}^{+0.106}$ \\\vspace{0.05cm}
 28 &  02:17:57.60 &  $-$05:08:44.9 &        6.595 &  25.64 $\pm$ 0.14 &  23.67 $\pm$ 0.05 &          $<${0.68} &  ${43.68}_{-0.02}^{+0.02}$ &  ${1.834}_{-0.004}^{+0.004}$ \\\vspace{0.05cm}
 29 &  10:00:58.01 &  $+$01:48:15.1 &        6.607 &  25.30 $\pm$ 0.09 &  23.59 $\pm$ 0.05 &          $<${0.66} &  ${43.85}_{-0.02}^{+0.02}$ &                          --- \\\vspace{0.05cm}
 30 &  02:39:11.86 &  $-$01:38:51.3 &        6.609 &  26.77 $\pm$ 0.36 &  24.65 $\pm$ 0.07 &          $<${0.63} &  ${43.45}_{-0.02}^{+0.02}$ &  ${1.737}_{-0.015}^{+0.013}$ \\\vspace{0.05cm}
 31 &  09:59:06.22 &  $+$01:59:51.9 &        6.612 &  25.77 $\pm$ 0.16 &  24.69 $\pm$ 0.13 &  ${0.82}\pm{0.38}$ &  ${43.46}_{-0.06}^{+0.05}$ &  ${1.742}_{-0.043}^{+0.032}$ \\\vspace{0.05cm}
 32 &  02:16:54.56 &  $-$04:55:57.0 &        6.618 &           $>$27.0 &  25.16 $\pm$ 0.15 &          $<${0.32} &  ${43.37}_{-0.06}^{+0.05}$ &  ${1.674}_{-0.055}^{+0.042}$ \\\vspace{0.05cm}
 33 &  02:40:14.71 &  $-$01:46:00.9 &        6.621 &  26.44 $\pm$ 0.28 &  24.82 $\pm$ 0.09 &          $<${0.72} &  ${43.59}_{-0.03}^{+0.03}$ &  ${1.811}_{-0.012}^{+0.010}$ \\\vspace{0.05cm}
 34 &  02:18:44.64 &  $-$04:36:36.1 &        6.622 &  26.41 $\pm$ 0.29 &  24.88 $\pm$ 0.11 &          $<${0.80} &  ${43.62}_{-0.04}^{+0.04}$ &  ${1.821}_{-0.015}^{+0.009}$ \\\vspace{0.05cm}
 35 &  02:17:55.56 &  $-$05:10:48.1 &        6.624 &           $>$27.0 &  25.22 $\pm$ 0.20 &          $<${0.55} &  ${43.49}_{-0.07}^{+0.07}$ &  ${1.763}_{-0.047}^{+0.036}$ \\\vspace{0.05cm}
 36 &  10:02:05.92 &  $+$02:14:05.4 &        6.625 &  27.00 $\pm$ 0.45 &  24.71 $\pm$ 0.08 &          $<${0.75} &  ${43.75}_{-0.03}^{+0.03}$ &                          --- \\
\enddata
\tablecomments{The limits listed in the table correspond to $2\sigma$ upper bounds. The redshift errors are smaller than 0.001. The three LAEs, No.~13, No.~29, and No.~36 (in the COSMOS1 and COSMOS3 pointings with bad alignment), are not given with $V_a$ to mark that they are not adopted in the LF calculation.}
\end{deluxetable}

\begin{figure*}
\centering
\includegraphics[angle=0, width=0.85\textwidth]{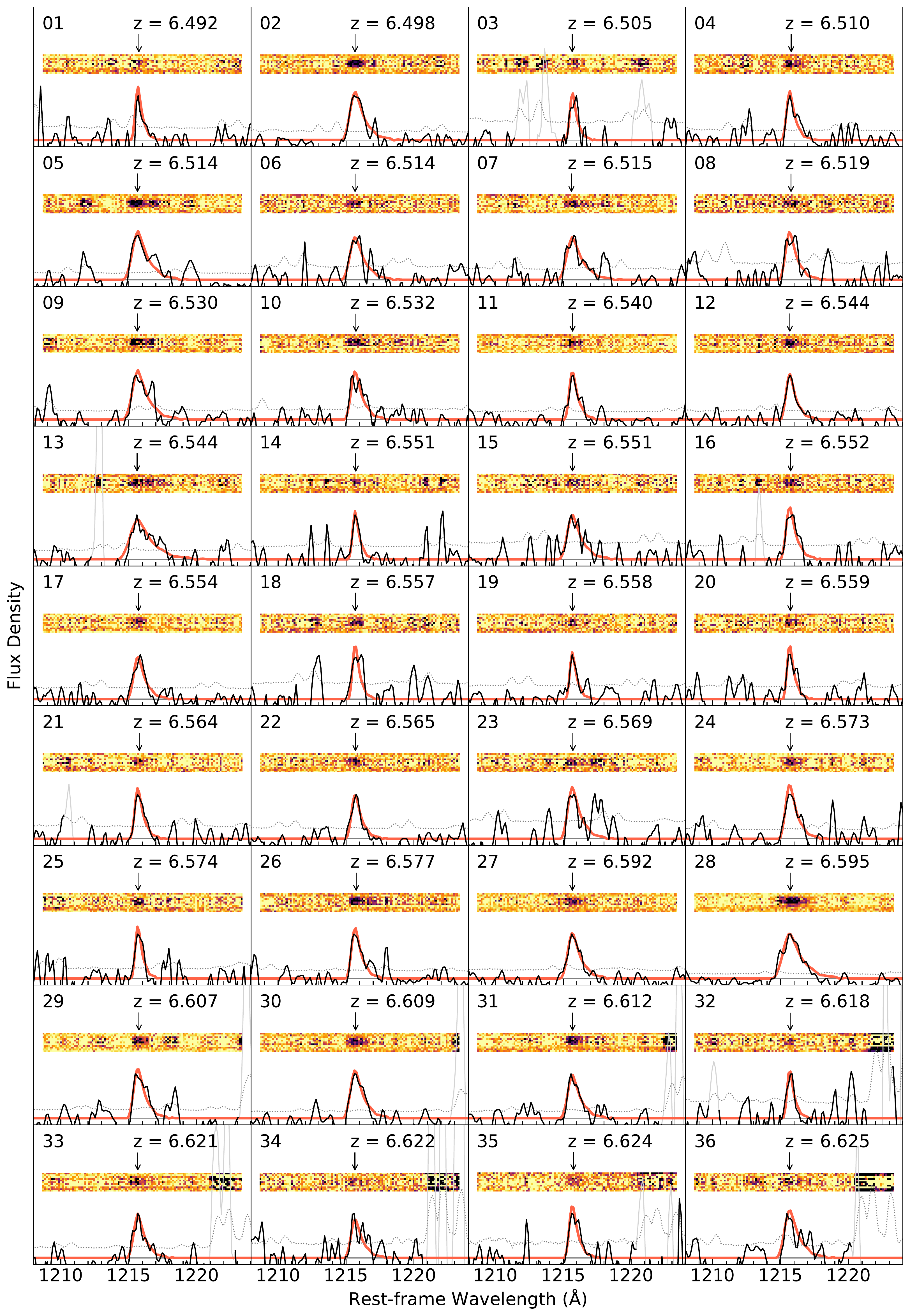}
\caption{M2FS 1D and 2D spectra in the rest frame of 36 LAEs in our $z\approx6.6$ sample. In each panel, the 1D spectrum have been smoothed with a Gaussian kernel (a $\sigma$ of one pixel is used) and its abnormal values are plotted in gray, while darker color represent higher flux in the 2D spectrum. The gray solid and dotted lines indicate zero and $1\sigma$ uncertainty level, respectively. The best-fit \lya\ line template is overplotted in red. The downward arrow points to the position of the \lya\ emission line. The source numbers correspond to those shown in Column 1 in Table \ref{sample}.
\label{glaes}}
\end{figure*}

\begin{figure*}
\centering
\includegraphics[angle=0, width=1.0\textwidth]{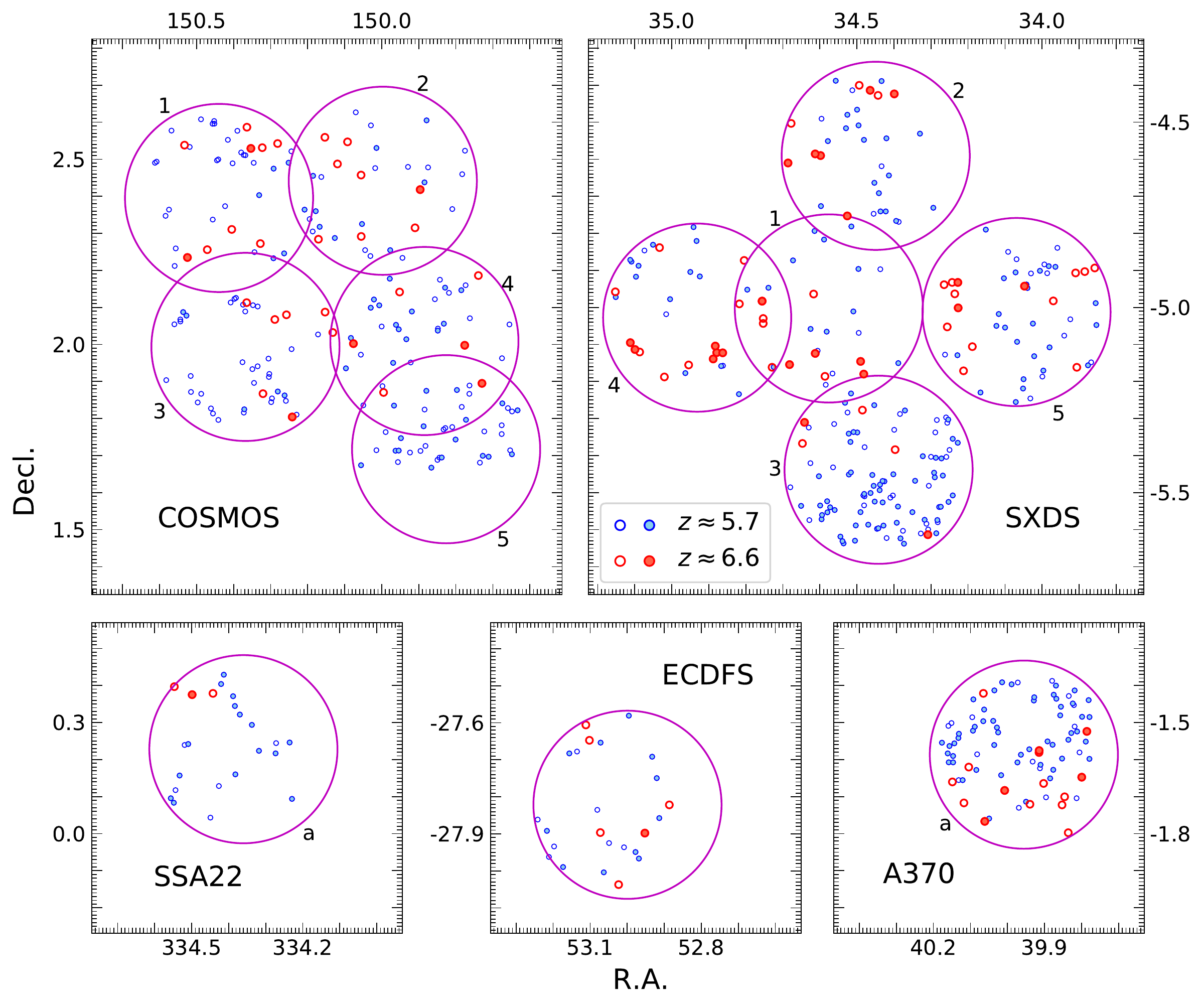}
\caption{The five deep fields observed by our M2FS survey. The big circles represent our M2FS pointings. All points represent the LAE targets observed by our M2FS survey. The filled points represent the spectroscopically confirmed LAEs. Two colors correspond to two redshifts of  $z\approx5.7$ (from \pn) and $z\approx6.6$ (this work).
\label{fields}}
\end{figure*}

\section{Spectroscopic Results}

\begin{figure}[t]
\epsscale{1.15}
\plotone{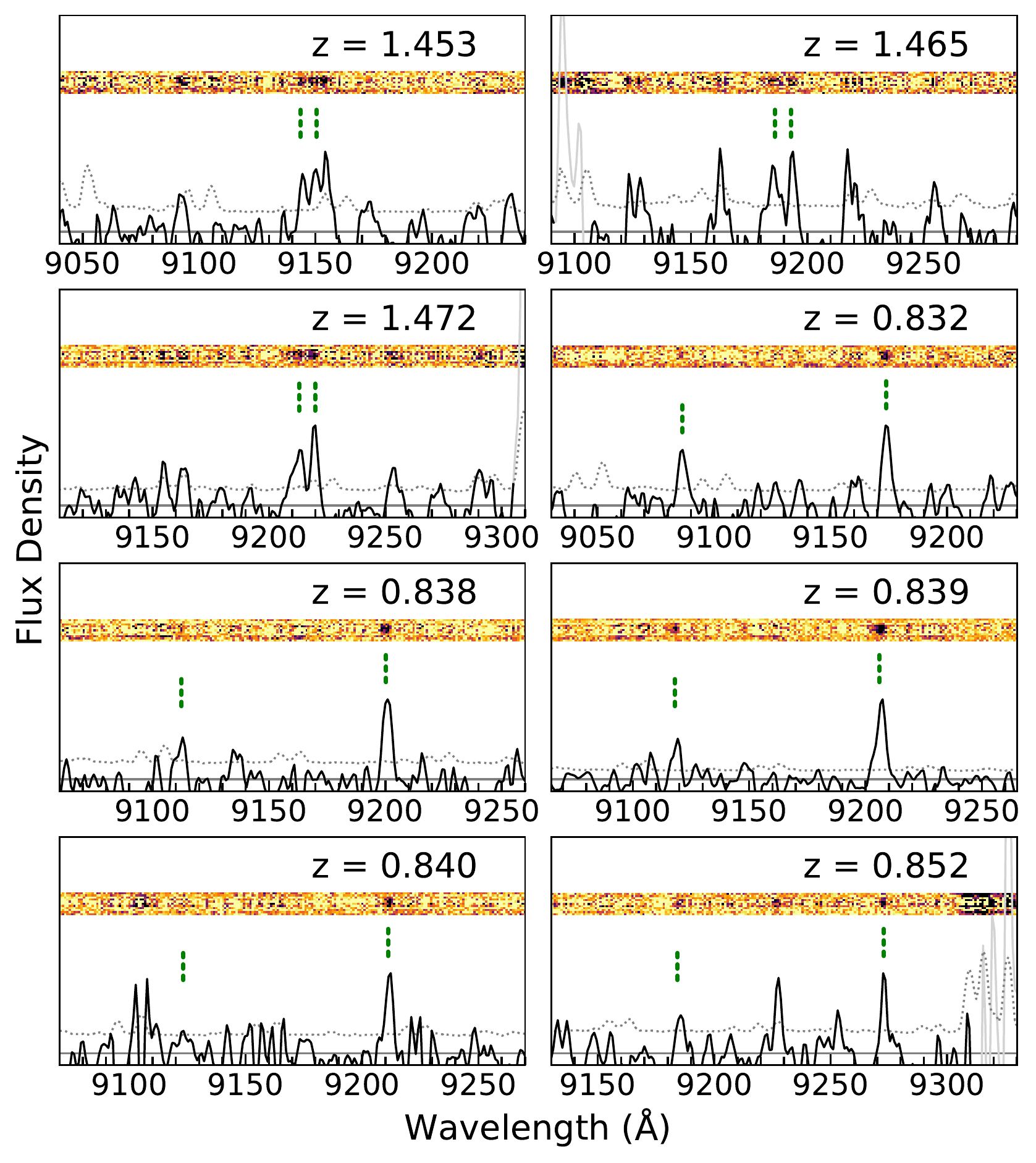}
\caption{M2FS 1D and 2D spectra of the eight low-redshift interlopers. Besides the same graphic elements as Figure \ref{glaes}, the vertical green dotted lines mark the positions of the [O\,{\sc ii}] and [O\,{\sc iii}] doublets.\label{lowz}}
\end{figure}

\subsection{\lya\ Identification}

We follow the approach in Section 3 of N20 to identify LAEs and reject contaminants. Our selection criteria generally ensure that an emission line detected in the expected wavelength range is the \lya\ line, based on the non-detection in the bands bluer than $R$ or $r'$. But occasionally it can be one of other emission lines from low-redshift interlopers, such as \oii, \hb, \oiii, or \ha\ lines. The \oii\ doublet and the \oiii\ line are common contaminant lines for high-redshift, narrowband-selected galaxies. Galaxies with strong [O\,{\sc ii}] emission can be very faint in the {\it BVR} images because there are no strong emission lines in the wavelength range between \lya\ and the doublet. Our resolving power of $\sim$2000 can nearly resolve the doublet to identify [O\,{\sc ii}]. \hb, \oiii, and \ha\ usually show symmetric and relatively narrow shapes. A high-redshift LAE generally has an asymmetric \lya\ emission line with a red wing due to internal interstellar medium (ISM) and circumgalactic medium (CGM) kinematics and IGM absorption \citep{santos04}. This characteristic feature serves as one criterion to identify \lya\ emission.

We identify \lya\ emission lines based on both 1D and 2D spectra. First, each 1D spectrum is smoothed with a Gaussian kernel (a $\sigma$ of one pixel is used). Then, we check if it has an emission line with S/N~$\gtrsim5$ in the expected wavelength range around 9200 \AA. This line signal needs to cover at least five contiguous pixels above $1\sigma$ noise level in the smoothed spectrum. We estimate its S/N by stacking the corresponding pixels in the original spectrum. Next, we visually inspect it in the individual and combined 2D spectra. \lya\ emission shows a diffuse pattern due to its asymmetry with an extended red wing. Spurious or unreliable detections can be easily removed, for example, a detection that is part of cosmic ray residuals in the 2D image or only shows up in one of the individual 2D images. We remove a strong line from a low-redshift galaxy by its two characteristics: a narrow profile with comparable width to the point spread function (PSF); and a compact shape without a tail redward in the 2D spectrum.

Finally, we spectroscopically confirmed 36 LAEs at $z\approx6.6$. They are summarized in Table \ref{sample}. Column 4 lists the spectroscopic redshifts measured from the \lya\ lines. Columns 5 and 6 show their photometry in $z'$, and NB921, respectively. Column 7 lists the estimated continuum flux density at the observed \lya\ wavelength. Column 8 lists the \lya\ luminosities. Column 9 shows their available volume $V_a$. Figure \ref{glaes} shows their 1D and 2D spectra. The 1D spectra are shown in arbitrary units for clarity. We can see that strong \lya\ emission lines usually show asymmetric line shapes due to the ISM, CGM kinematics and IGM absorption. Figure \ref{fields} illustrates the positions of the LAE targets of our survey in the five fields, including the observed candidates (all points) and the
confirmed LAEs (filled points).

\subsection{Confirmation Rate}

From the candidates, we found eight low-redshift interlopers (Figure \ref{lowz}). A small fraction of the remaining targets show plausible line features with low S/N values. We do not include them in our LAE sample because they do not satisfy the line identification criteria. The rest of the targets do not show obvious emission features in the spectra. We observed LAE candidates at $z\approx5.7$ and $z\approx6.6$ simultaneously, but the sensitivity for $z\approx6.6$ LAEs is lower, due to the decline of the system efficiency towards longer wavelength in the red end.

\begin{figure}[t]
\epsscale{1.15}
\plotone{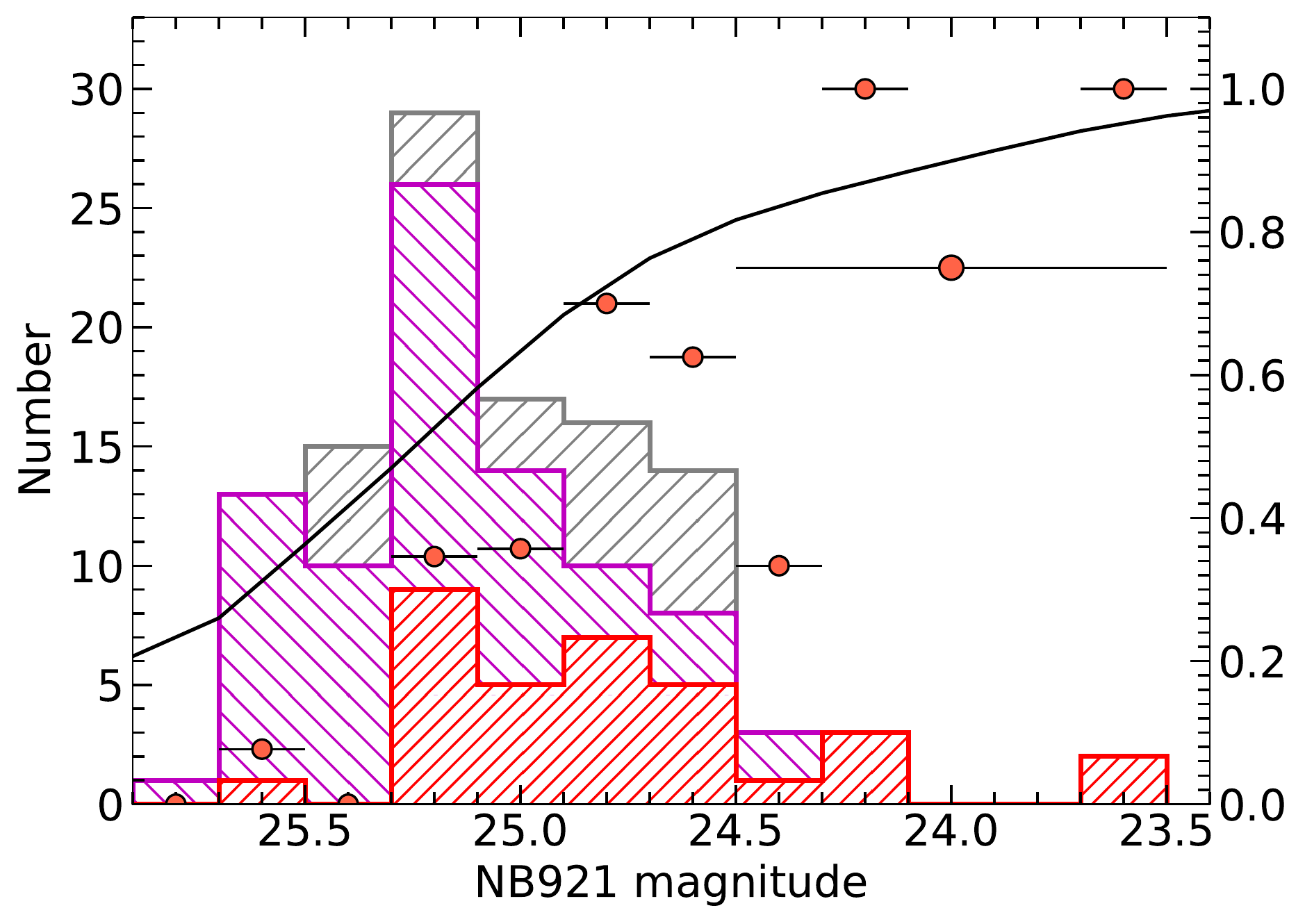}
\caption{Distribution of the NB921 magnitudes for all LAE candidates (grey histogram), observed targets (magenta histogram), and spectroscopically confirmed LAEs (red histogram). The red points denote the spectroscopic confirmation rate that is defined as the ratio of the confirmed number to the targeted number in each magnitude bin. That with a longer error bar indicates an average value for the four brightest magnitude bins. The curve is the confirmation rate from our simulations. 
\label{cfr}}
\end{figure}

We plot the spectroscopic confirmation rate as a function of NB921 magnitude in Figure \ref{cfr}. The rate is defined as the ratio of the numbers of the confirmed LAEs to the targeted candidates in each magnitude bin. Note that it is not fully equivalent to the \lya-identification fraction in Section 4.3. The measured rate increases toward bright magnitude as a whole, but shows a large scatter due to a small number of bright candidates. Besides those from observations (the red points), we also plot a simulated result by a solid curve. The simulated curve is obtained using the datasets for measuring the completeness of the candidate selection (Section 4.2) and \lya-line identification (Section 4.4). It is a monotonic function of the magnitude because more weak \lya\ lines (low identification fraction) may exist in fainter magnitude bins. The curve is also overall higher than the red points. This is because we did not add contamination in our simulations. At magnitude of ${<}24.5$, the mean confirmation rate reaches 75\% (6/8) (the red symbol with the error bar). It reveals a non-negligible contamination rate in the photometric sample. Such phenomenon was also reported before. For example, a robust selection criteria (a NB limit of 23.5 and a color cut of 1.3) just give a LAE confirmation rate of $\sim$70\% \citep{songaila18, taylor20, taylor21}. 

\subsection{\lya\ Redshift and Luminosity}

We calculate LAE redshifts using the composite template of the \lya\ profile (at $z\approx5.7$) from \pn. The \lya\ line template has the central wavelength of $\lambda_{\rm Ly\alpha}=1215.67$ \AA, and is scaled so that its peak value is 1 (arbitrary units). For each LAE, we use the wavelength of the \lya\ line peak to estimate its initial redshift.
From the composite template, we generate a set of model spectra with a grid of peak value, line width, and redshift. The peak value, by scaling the composite line, is from 0.8 to 1.2 with a step size of 0.01. The line width, by shrinking and expanding the composite line, is from 0.5 to 2.0 times the original width with a step size of 0.1 (times the original width). The redshift value varies within the initial redshift $\pm0.004$ (a spectroscopic resolving power of ${\sim}2000$ corresponds to $(1+z)/2000\approx0.004$) with a step size of 0.0001. Finally, we fit the \lya\ line of the LAE using the above model spectra and find the best fit. The wavelength range used in the fitting process is $(1+z)$ $\times$ [1215.67--1,\ 1215.67+3] \AA.

\begin{figure}[t]
\epsscale{1.15}
\plotone{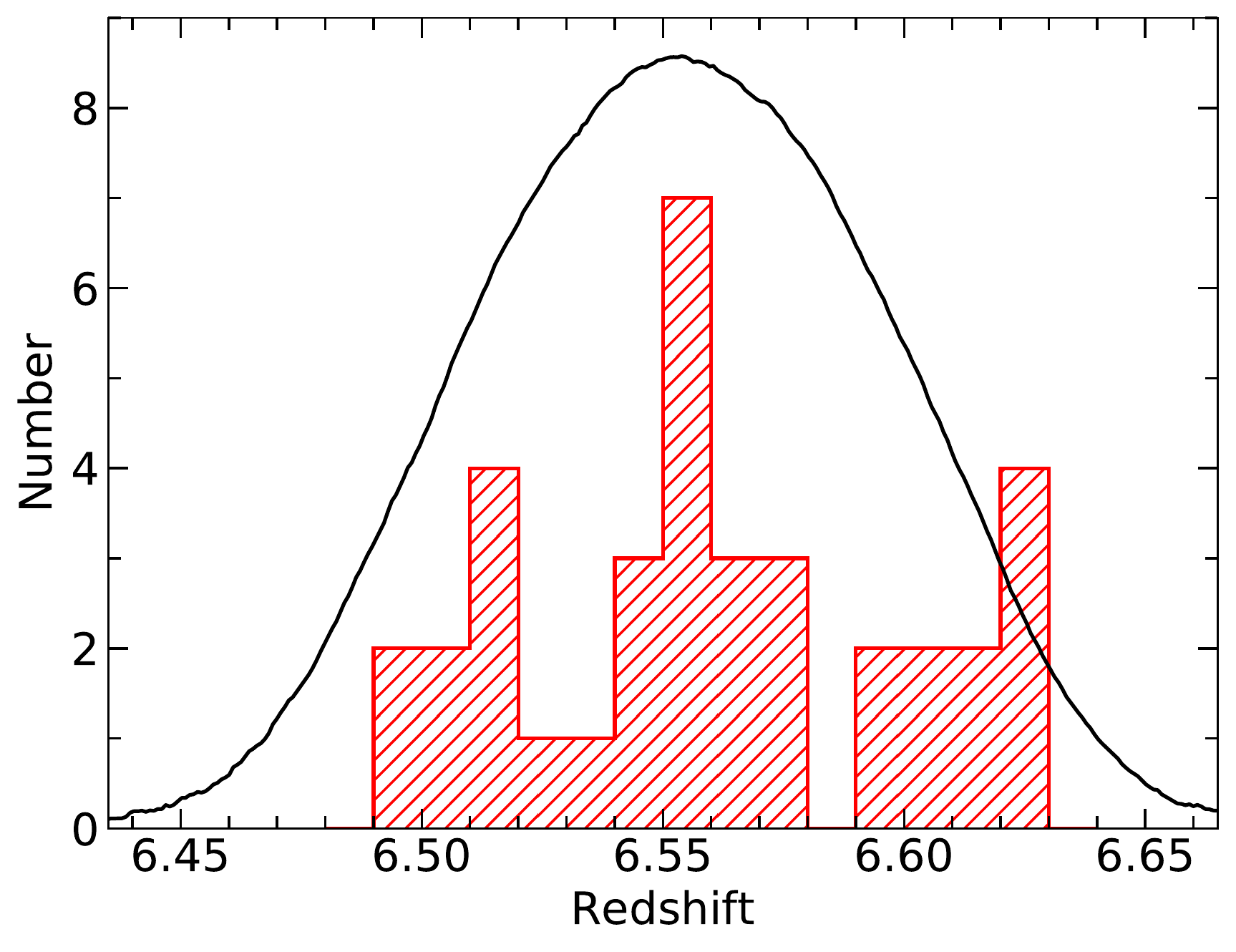}
\caption{Redshift distribution of the 36 LAEs. The NB921 filter transmission curve is over-plotted and scaled for clarity.
\label{zl}}
\end{figure}

Figure \ref{zl} shows the redshift distribution of the 36 LAEs at $z\approx6.6$. The number excess near the higher-redshift edge is caused by very luminous LAEs that contributed to the bright-end bump in the LF. \citet{ning20} reported a clear offset between the redshift distribution of 260 $z\approx5.7$ LAEs and the NB816 filter transmission curve. Here we do not see such an offset. This is likely because of the much smaller number of our $z\approx6.6$ LAEs. More LAEs are needed to improve the measurement of this redshift distribution.

We estimate the \lya\ line flux using the narrowband (NB921) and broadband ($z'$) photometry. We use a model spectrum, which is the sum of a \lya\ emission line and a power-law UV continuum with a slope $\beta$,
\begin{eqnarray}
   f_{\lambda} = f_{\rm Ly\alpha}\times{P}(\lambda) + f_{\rm cont}\times\lambda^{\beta},
\label{flambda}
\end{eqnarray} where $f_{\rm Ly\alpha}$ and $f_{\rm cont}$ in units of $\rm erg\ s^{-1}\ cm^{-2}$ \AA$^{-1}$ are scale factors of the \lya\ line flux and the UV continuum flux, respectively, and $P(\lambda)$ is the dimensionless line profile of our template that is assigned with the three fitting parameters (peak value, line width, and redshift) and redshifted to the observed frame for each LAE. We adopt an average $\beta=-2.3$ from a sample of spectroscopically confirmed LAEs at $z\gtrsim5.7$ by \citet{jiang13a}, because the UV slope $\beta$ can not be determined for individual LAEs. We consider the IGM absorption of continuum emission blueward of \lya\ in the model spectrum \citep{madau95}. For each LAE with $z'\ge2\sigma$ detection, we match its model spectrum to both NB921 and $z'$ magnitudes to determine $f_{\rm Ly\alpha}$ and $f_{\rm cont}$. For each LAE with $z'<2\sigma$ detection, we match its model spectrum to the NB921 magnitude and $2\sigma$ upper limit in $z'$ band. In a case with a very weak detection in the $z'$, our calculation may produce negative $f_{\rm cont}$. In this case, the continuum flux is negligible and the \lya\ emission is very strong, so we assume $f_{\rm cont}=0$. After obtaining $f_{\rm Ly\alpha}$, we measure the \lya\ line flux and then calculate the \lya\ luminosity using the cosmological luminosity distance derived from its spectroscopic redshift. To incorporate the errors of photometric data, we simulate the probability distribution of \lya\ luminosity based on the input NB921 and $z'$ magnitudes and their errors. The result is shown in Column 8 of Table \ref{sample} with their $1\sigma$ errors.

\begin{figure}[t]
\epsscale{1.15}
\plotone{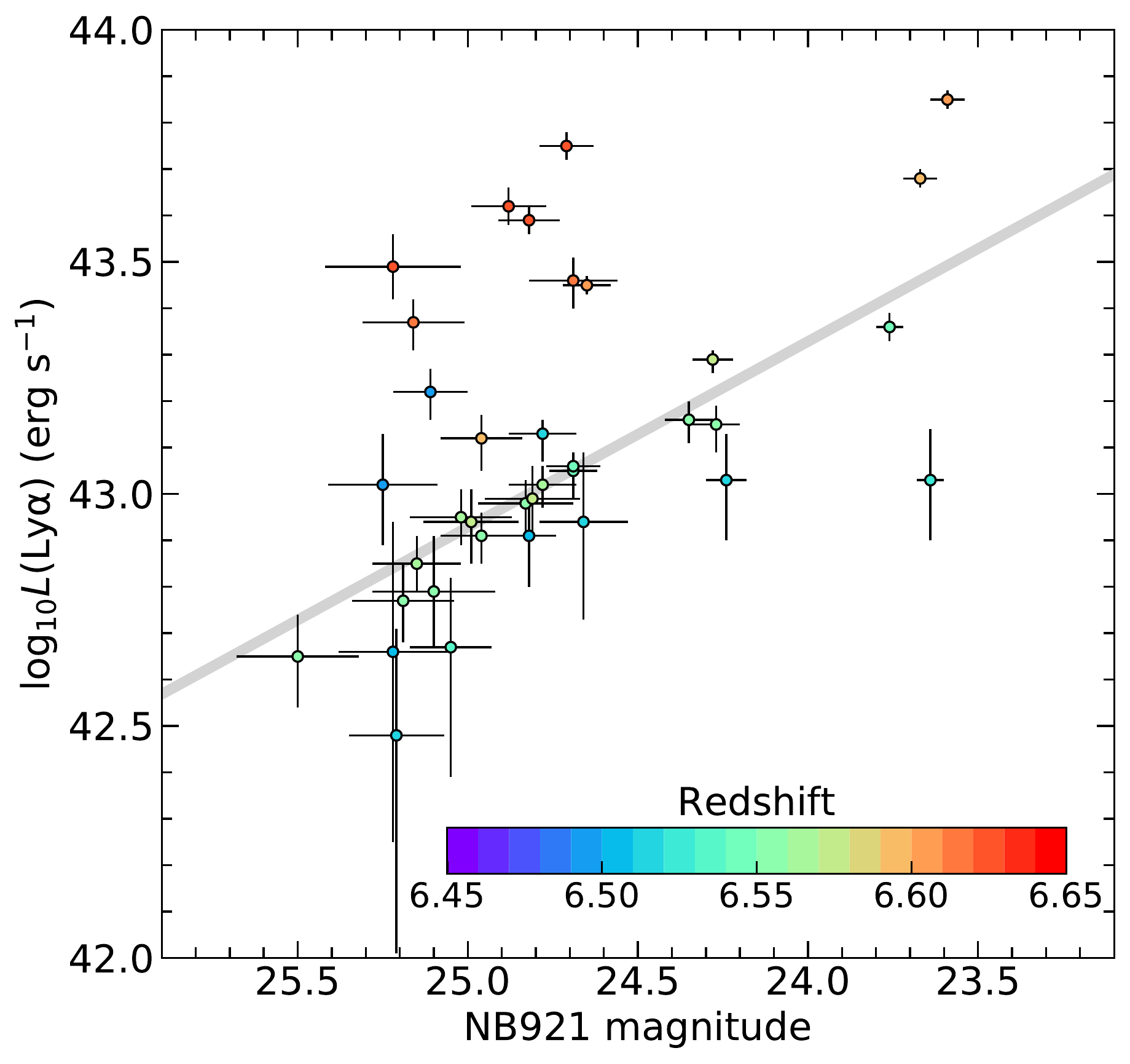}
\caption{\lya\ luminosity vs. NB921 magnitude for our LAE sample. Each symbol is color-coded by redshift with $1\sigma$ error bars. The gray line is calculated assuming LAEs solely have \lya\ line at the center of NB921 filter transmission.
\label{mag2l}}
\end{figure}

Figure \ref{mag2l} shows the \lya\ luminosities and narrowband magnitudes of the 36 LAEs at $z\approx6.6$. Our LAE sample spans a luminosity range of ${\sim} 3\times10^{42} {-} 7\times10^{43}$ erg~s$^{-1}$. Previous studies using photometric samples usually assumed that the \lya\ line locates at the central wavelength (or effective wavelength) of the narrowband filter due to the lack of the redshift information \citep[e.g.,][]{ouchi08, ouchi10, zheng17, shibuya18a, hu19}. We do the same calculation for comparison and plot the result as the gray line in Figure \ref{mag2l}. We can see that some LAEs in our sample are located close to the gray line because their redshifts correspond to wavelengths near the filter center. Some LAEs are below the line due to the non-negligible continuum indicated by the $z'$-band detections. It is remarkable that a fraction of LAEs, particularly very luminous LAEs, largely deviate from the line. The reason is that the filter curve does not have a perfect top-hat profile and thus the \lya\ redshift also determines the narrowband detection, as explained as follows. In an imaging survey, a  luminous LAE at the filter edge (with a low transmission) can be faint, as shown by Figure \ref{mag2l}, and its \lya\ luminosity can be underestimated by the method used for photometric samples. The filter-profile correction is thus necessary for the statistical analysis of LAE luminosities. We will further address this correction in the measurement of sample completeness (see Section 4.1.2).

\subsection{\lya\ Luminosity vs. Line Width}

The line profile of \lya\ emission can probe the dynamical structures of \lya\ emitting galaxies \citep{dijkstra14}. We have investigated the relation between the line width and luminosity of the \lya\ emission from the sample of LAEs at $z\approx5.7$ in \pn. We adopt the same approach for the $z\approx6.6$ sample. According to \lya\ luminosities, we divide the $z\approx6.6$ sample into three subsamples. For each subsample, we stack \lya\ spectra in the rest frame and measure the FWHMs. Our division ensures that the three stacked \lya\ lines have similar S/Ns that are as high as possible for measuring line widths. We have excluded eight LAEs with low S/Ns.

\begin{figure}[t]
\epsscale{1.15}
\plotone{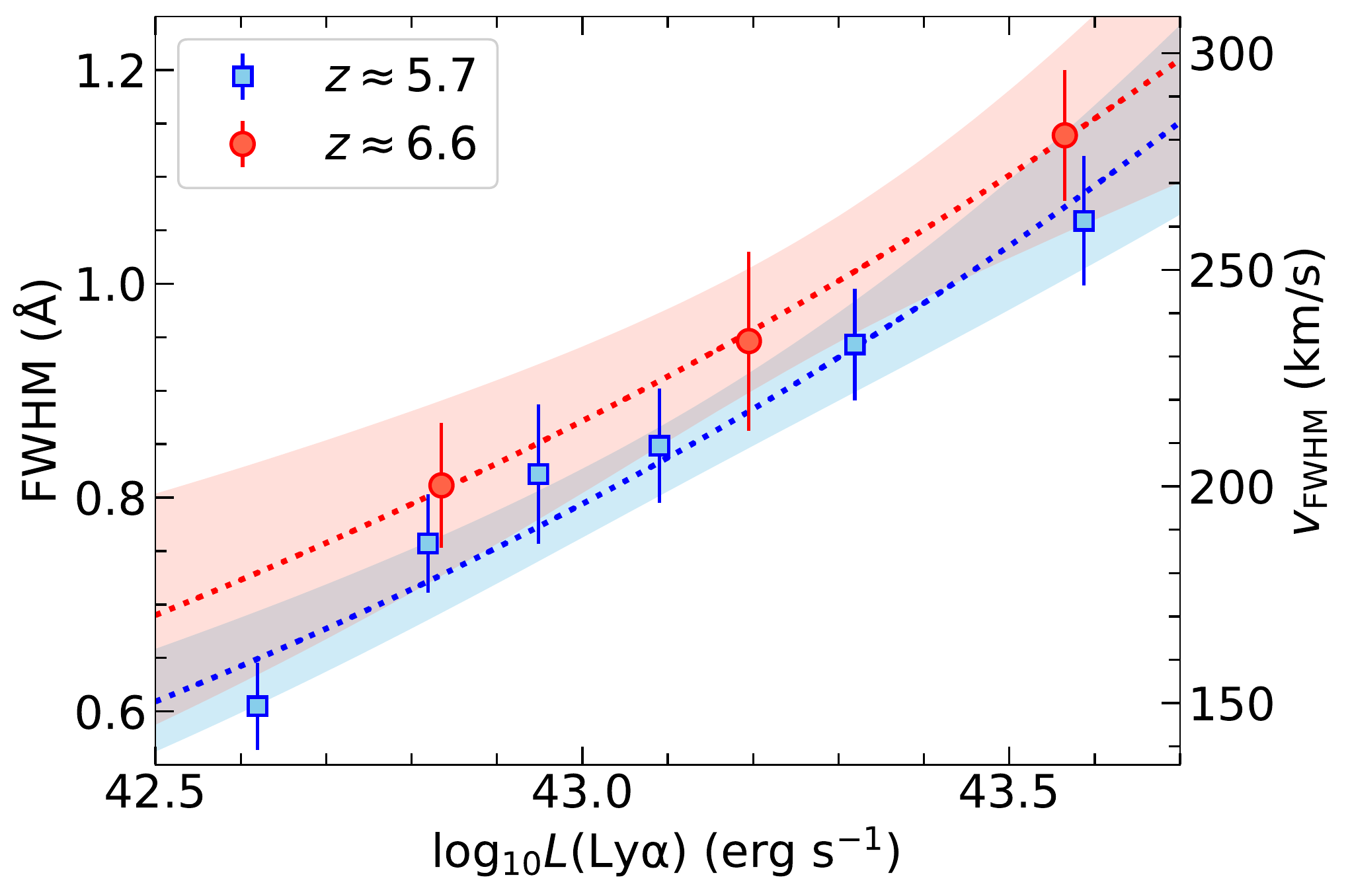}
\caption{The intrinsic line width of the stacked \lya\ line as a function of \lya\ luminosity. Each data point has the measurement error of the stacked spectrum. The dotted lines represent the best-fit power-law functions while the corresponding colored shades cover $1\sigma$ error region of the fitting.
\label{stack}}
\end{figure}

The intrinsic \lya\ line widths in the rest frame are shown in Figure \ref{stack}. The \lya\ line width increases toward higher luminosities for both $z\approx5.7$ and $z\approx6.6$ LAEs. By fitting a power-law relation of ${\rm FWHM}=A L^{\gamma}$, we obtain $\gamma=0.20\pm0.05$ and log$_{10}A=-8.8\pm2.2$ (red dotted line). The best fit for $z\approx5.7$ has $\gamma=0.23\pm0.03$  and log$_{10}A=-10.0\pm1.4$ (blue dotted line). The two power-law relations agree with each other within $1\sigma$. They are consistent with the observational results of \citet{hu10} and the previous simulation results \citep[e.g.,][]{weinberger18, sadoun19}. In general, more luminous LAEs emit broader \lya\ line because their more massive host halos possess higher neutral hydrogen column densities and higher gas velocities in the ISM and CGM.

\section{\lya\ luminosity function}

\subsection{Sample Completeness}

To calculate the \lya\ LF, we measure the sample completeness. Sample incompleteness originates from four sources: object detections in imaging data, candidates selection, spectroscopic observations, and \lya-line identification. The following subsections provide details about completeness correction from the four sources for our sample.

\subsubsection{Detection Completeness}

\begin{figure*}
\centering
\includegraphics[angle=0, width=0.9\textwidth]{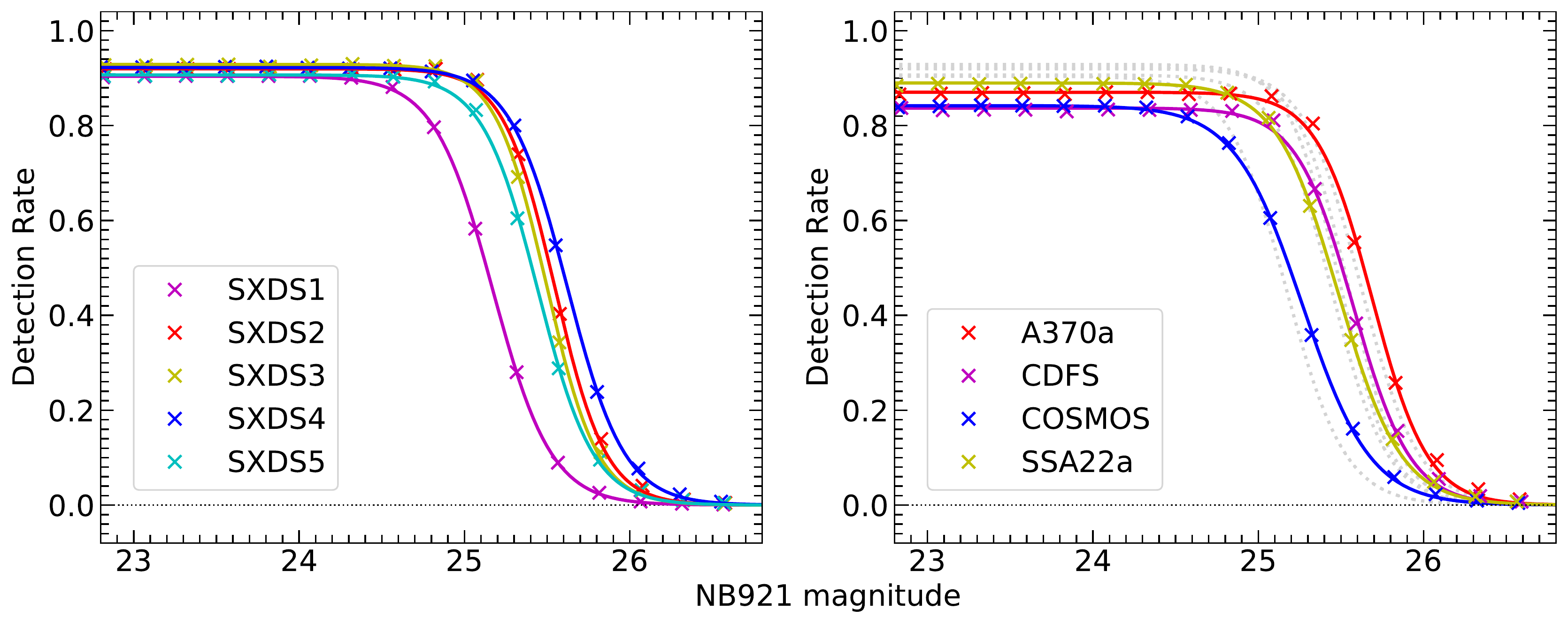}
\caption{Detection rates in NB921. The equation (\ref{fermidirac}) is used to fit the measured detection rates for each field. The colored best-fit curves in the left panel are also shown in the right panel, but in gray dotted, for comparison. See the text for details.
\label{ic1}}
\end{figure*}

The first incompleteness originates from the object detections in imaging data. We implement Monte Carlo simulations to measure the completeness fraction. LAEs are reported with diffuse \lya\ halos \citep{wisotzki16, leclercq17, xue17, wu20}. The potential, extended morphology affects object detection completeness and thus \lya\ LF. We properly create mock LAE sources for each field and randomly scatter them in the images.

We first consider the point-spread function (PSF), a key quantity to determine the detection limit for an image. In this work, we have nine fields with different PSFs. For each field, we obtain its PSF by co-adding more than 100 bright (not saturated) stars in the NB921 image. Next, we stack ${\gtrsim}200$ spectroscopically confirmed LAEs at $z\approx5.7$ in the SDF \citep[Subaru Deep Field;][]{kashikawa04}, A370, and SXDS \citep{ning20, wu20}. The three fields have an image quality of ${\gtrsim}0.5\arcsec$ in the NB816 images. The stacked LAE has a PSF FWHM $\approx0.6\arcsec$. Since the intrinsic sizes of $z\ge6$ galaxies ($<0.2\arcsec$) are typically much smaller than the PSF sizes, we use this $z\approx5.7$ stacked LAE to represent a typical morphology of $z\approx6.6$ LAEs. We then compare this PSF with the PSF of each NB921 image to derive the matching convolution kernel, and convolve the matching kernel with the stacked LAE to obtain the mock LAE in each field.

For each field, we simulate mock LAEs and insert them into the NB921 image. At a given magnitude, $\gtrsim$100,000 mock LAEs per deg$^2$ are randomly inserted into the NB921 image. The minimum separation between any two LAEs is 25 pixels. We only consider the mock LAEs in the region covered by the corresponding M2FS pointing. We then run SExtractor \citep{ber96} to detect the mock sources. We also require that they are in the clean region of the corresponding $z'$-band image used for the color-cut selection. The detection completeness is the fraction of the inserted mock sources that are detected. The results are plotted in Figure \ref{ic1}. Colored curves are the best fits to the measured detection rates. The curves have a functional form, 
\begin{eqnarray}
f(m) = A\left[1+{\rm exp}\left(\frac{m-m_0}{\Delta m}\right)\right]^{-1},
\label{fermidirac}
\end{eqnarray} where $m_0$ and $\Delta m$ are two magnitude parameters and $A$ is a normalization factor. It performs well for the detection rate decreasing as the magnitude goes deeper, as shown in Figure \ref{ic1}. In the following LF calculation, we adopt these best-fit curves to correct the sample completeness.

\subsubsection{Selection Completeness}

\begin{figure*}
\centering
\includegraphics[angle=0, width=0.95\textwidth]{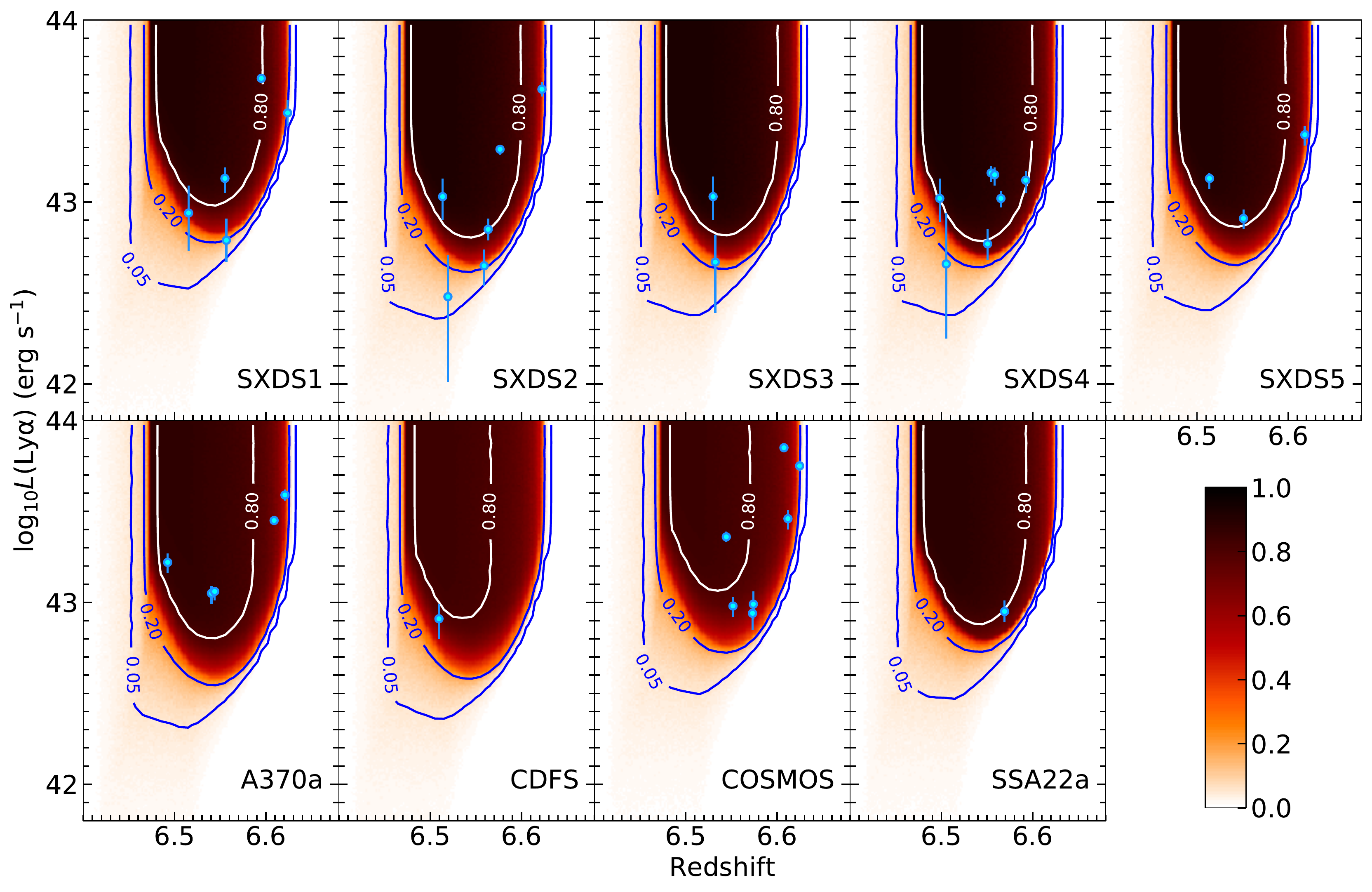}
\caption{The completeness fraction as a function of ${\rm log}_{10}L({\rm Ly\alpha})$ and $z$ for the imaging detection and our selection criteria. In each panel, three contours represent selection probabilities of 5\%, 20\%, and 80\%, respectively. The cyan symbols with error bars represent the LAEs in our sample.
\label{ic2}}
\end{figure*}

The second incompleteness originates from candidate selection. An LAE has a probability to be selected by the color-magnitude criteria. We estimate the completeness using simulations. We simulate mock LAE spectra using the model (\ref{flambda}) with $\beta=-2.3$, an average UV-slope value of a large sample of spectroscopically confirmed galaxies at $z\geq6$ \citep{jiang13a}, and $P(\lambda)$ adopted by the dimensionless composite \lya\ profile at $z\approx5.7$ from \pn. The \lya\ equivalent width (EW) is assumed to have an exponential distribution, 
\begin{eqnarray}
  \frac{dN}{d\rm EW} \propto {\rm exp}\left(-\frac{\rm EW}{W_0}\right),
  \label{ewdist}
\end{eqnarray} where we adopt $W_0=168$ \AA\ from \citet{shibuya18a}. The EW scale length is consistent with previous estimations at high redshift in the literature \citep[e.g.,][]{zheng14, hashimoto17b}. We construct a grid of ${\rm log}_{10}L({\rm Ly\alpha})$ and $z$. The \lya\ luminosity $L({\rm Ly\alpha})$ is in the logarithmic range of $41.8{-}44.0$ with a step size of 0.01. The redshift $z$ is in the range of $6.40{-}6.68$, corresponding to the NB921 bandpass range, with a step size of 0.002. In each mesh of $\left[{\rm log}_{10}L({\rm Ly\alpha}), z\right]$, 2,000 mock LAE spectra are simulated with their NB921 and $z'$-band magnitudes. We produce photometric errors following the magnitude-error relations measured from real images of each field. The selection completeness for this $\left[{\rm log}_{10}L({\rm Ly\alpha}), z\right]$ mesh is the fraction of the mock LAEs passing our selection criteria (see Section 2.2). Figure \ref{ic2} shows the results including the detection completeness for all fields.

\subsubsection{Observation and Identification Completeness}

The third incompleteness originates from spectroscopic observations. It is the fraction of photometrically selected LAE candidates that were spectroscopically observed. For each field, we only consider the region covered by the M2FS pointing(s). As we mentioned in Section 2.3 of \pn, COSMOS1 and COSMOS3 suffered serious alignment problems. In the following LF calculation, we exclude the three LAEs (No.~13, No.~29, and No.~36) in the two pointings and use the coverage area of the remaining COSMOS (2, 4, and 5) pointings. The three LAEs are very luminous in terms of \lya. No.~29 is known as CR7 \citep{sobral15}. They can be detected due to their strong emission in spectra, even without good pointing alignments. In Figure \ref{glaes}, \lya\ lines of No.~29 and No.~36 do not have enough high S/N like No.~28, the third most luminous LAE in our sample \citep[known as Himiko;][]{ouchi13}. One observed target in SXDS5 does not have output spectra. The reason is unclear. It was previously confirmed with a \lya\ line with a \lya\ luminosity of $10^{42.95\pm0.14}$ erg~s$^{-1}$ \citep[NB921-C-50823;][]{ouchi10}. We include it in our LF calculation with the same completeness correction. We have 33 LAEs in our LF calculation.

\begin{figure}[t]
\epsscale{1.15}
\plotone{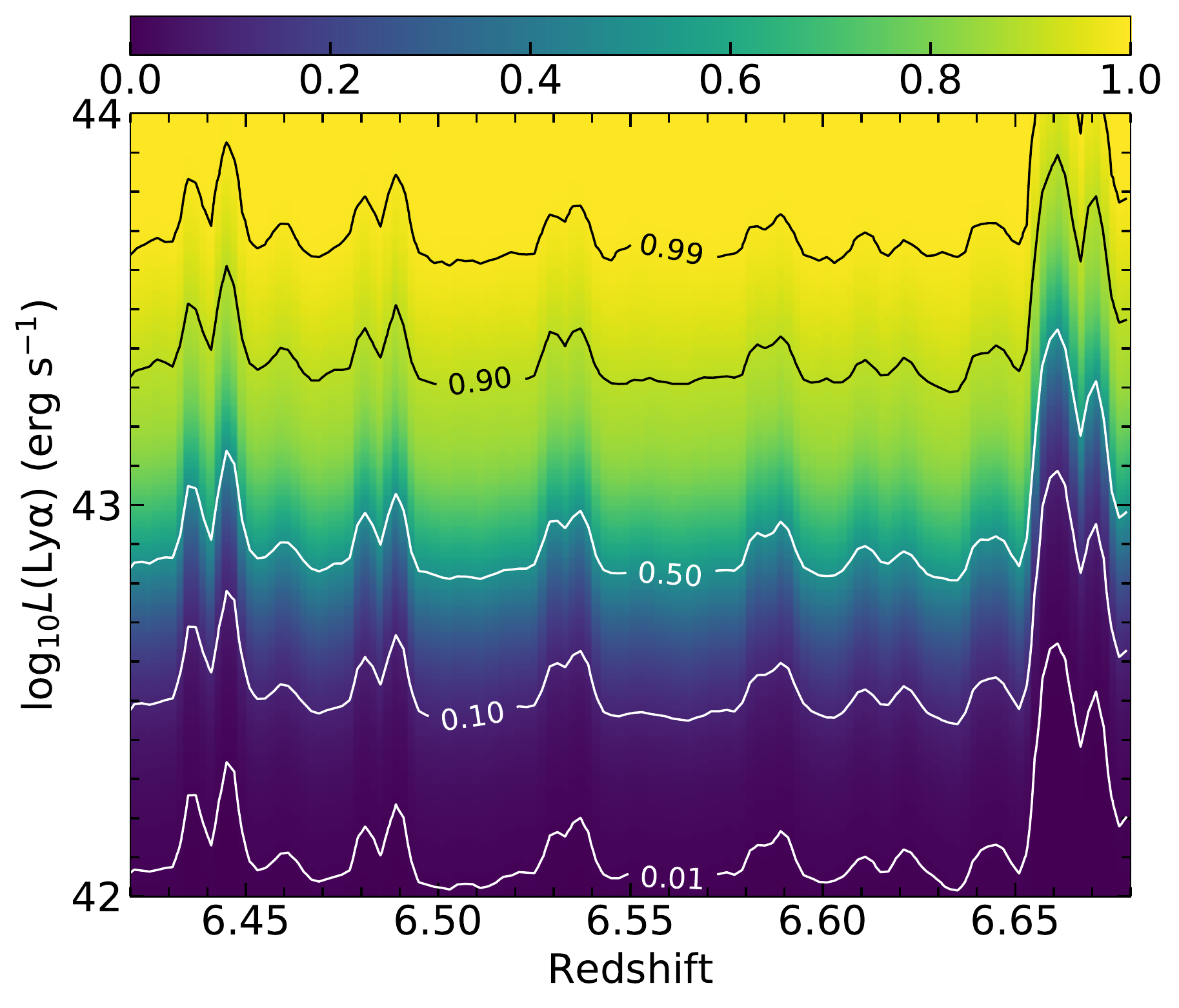}
\caption{\lya-identification completeness of our spectroscopic survey as a function of redshift and luminosity. The five contour lines denote the detection fractions of 1\%, 10\%, 50\%, 90\%, and 99\%. The hump-like features in the contours correspond to the locations of the (weak) skylines.
\label{ic3}}
\end{figure}

The fourth completeness originates from identifying \lya\ line in spectra. Here we run simulations based on the best-fit model \lya\ lines to give an estimation of the correction functions over all survey fields for the \lya-identification completeness. In our sample, each LAE has a best-fit line profile (see Section 3.2) corresponding to a redshift and \lya\ luminosity. For each of them, we shift its redshift and scale its luminosity to generate model \lya\ lines. We construct the same grid of ${\rm log}_{10}L({\rm Ly\alpha})$ and $z$ as described in Section 4.2. In each $\left[{\rm log}_{10}L({\rm Ly\alpha}), z\right]$ mesh, the model \lya\ line is inserted 1,000 times into the spectrum simulated by the noise. We calculate the fraction of successful identifications with the same approach as described in Section 3.1. Each LAE then has an identification completeness function of \lya\ redshift and luminosity. For each field, we obtain a mean function by averaging all LAEs therein. Equivalently, more than 10,000 model \lya\ lines are inserted into each $\left[{\rm log}_{10}L({\rm Ly\alpha}), z\right]$ mesh. Figure \ref{ic3} illustrates the mean completeness function of all fields. The completeness fraction is about half at $10^{42.9}$ erg~s$^{-1}$ and decreases to ${\sim}1\%$ at $10^{42}$ erg~s$^{-1}$. For each \lya\ luminosity, the fraction is smaller at the skyline locations, exhibiting the hump-like features in the contours.

We cross-check the line-identification completeness by comparing previous studies. \citet{ouchi10} has 16 spectroscopically confirmed LAEs at $z\approx6.6$ in the SXDS field. Three of them are not our candidates because they are relatively faint and not selected by our criteria. Another one is the observed target without output spectra as we mention above. In the remaining 12 LAEs that we observed, we confirm and match 7 LAEs at $z\approx6.6$. For the other 5 galaxies, we did not detect robust (S/N~$>5$) emission lines in their spectra. The ratio $7/12$ is consistent with the measured completeness fraction of ${\sim}0.6$ at ${\lesssim}10^{43.0}$ erg~s$^{-1}$ (their median \lya\ luminosity). We also match two LAEs among three targets confirmed by \citet{harikane19}. The two matched ones have \lya\ luminosities of ${\gtrsim}10^{43.0}$ erg~s$^{-1}$, higher than the unmatched one (${\lesssim}10^{42.8}$ erg~s$^{-1}$) with a smaller completeness fraction of $<0.5$. Other confirmed LAEs in \citet{harikane19} are not our candidates because they included fainter LAE samples selected with Subaru Suprime-Cam images.

\subsection{$1/V_a$ Estimate and Schechter-function Fitting}

We use the nonparametric method, the $1/V_a$ estimator \citep[e.g.,][]{ab80}, to derive the binned \lya\ LF at $z\approx6.6$ from our LAE sample. The available volume $V_a$ is defined as the comoving volume to discover a galaxy (LAE) of luminosity $L$ and redshift $z$. Here, $V_a$ is weighted by the completeness function if it exists. $V_a$ with given $L$ and $z$ in the bin $\Delta z$ is 
\begin{eqnarray}
V_a = \int_{\Delta z} p(L, z) \frac{dV}{dz} \,dz, 
\label{va0}
\end{eqnarray} where $p(L, z)$ is the completeness fraction as a function of luminosity $L$ and redshift $z$, and $\Delta z$ corresponds the redshift range determined by the NB921 filter. The LF is estimated as 
\begin{eqnarray}
\phi({\rm log}L) = \frac{1}{\Delta {\rm log}L}\sum_i \frac{1}{V_a^i},
\end{eqnarray} where the available volume $V_a^i$ is calculated for each galaxy, and the sum is over all galaxies in the bin $\Delta {\rm log}L$. The corresponding statistical uncertainty is given by
\begin{eqnarray}
\sigma[\phi({\rm log}L)] = \frac{1}{\Delta {\rm log}L}\left[\sum_i \left(\frac{1}{V_a^i}\right)^2\right]^\frac{1}{2}.
\end{eqnarray}

\begin{deluxetable}{ccc}[t]
\tablecaption{Binned Data of the \lya\ LF at $z\approx6.6$
\label{binlf}}
\tablehead{
\colhead{log$_{10}L$} & \colhead{N} &  \colhead{log$_{10}\phi$}\\
\colhead{(erg~s$^{-1}$)} & \colhead{($\Delta {\rm log}_{10}L=0.15$)} & \colhead{(cMpc$^{-3}\ [\Delta {\rm log}_{10}L]^{-1}$)}}
\startdata
 $42.65$ &   $3$ &  $-3.53\pm0.25$ \\
 $42.80$ &   $3$ &  $-4.21\pm0.26$ \\
 $42.95$ &  $10$ &  $-4.09\pm0.14$ \\
 $43.10$ &   $9$ &  $-4.31\pm0.15$ \\
 $43.25$ &   $2$ &  $-5.07\pm0.31$ \\
 $43.40$ &   $3$ &  $-4.93\pm0.25$ \\
 $43.55$ &   $3$ &  $-4.95\pm0.25$ \\
 $43.70$ &   $1$ &  $-5.44\pm0.43$ \\
\enddata
\end{deluxetable}

\begin{figure}[t]
\epsscale{1.15}
\plotone{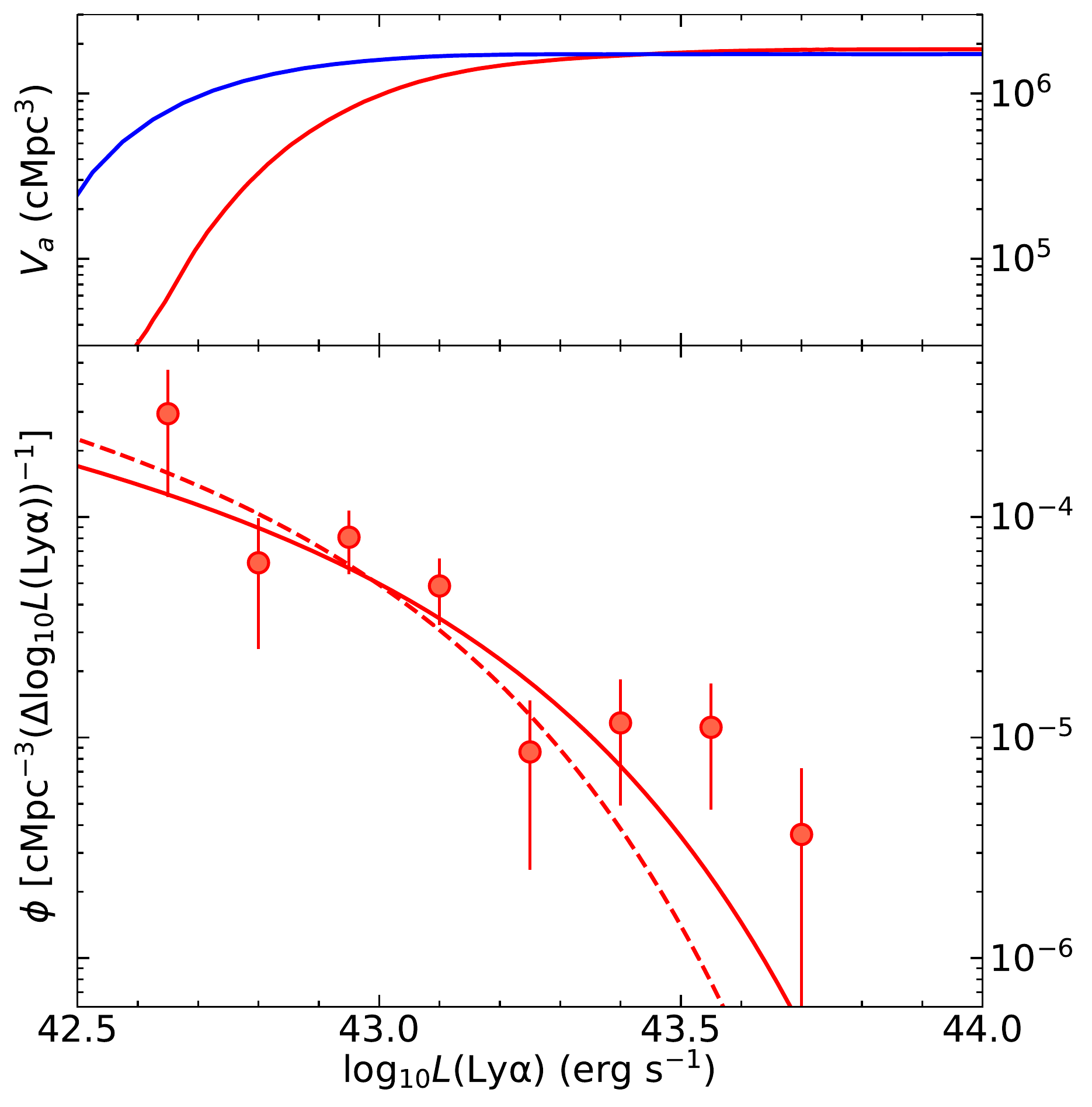}
\caption{Upper panel: the available volume $V_a$ as a function of \lya\ luminosity for our survey of LAEs at $z\approx6.6$ (red) and $z\approx5.7$ (blue). Lower panel: the $z\approx6.6$ binned \lya\ LFs and the Schechter functions fitted with and without the three brightest bins ($\alpha=-1.5$).
\label{valf}}
\end{figure}

In this work, our survey have nine fields with different sample completeness. We calculate $V_a$ by integrating over the solid angle $\Omega$, 
\begin{eqnarray}
V_a = \int_{\Delta\Omega} \int_{\Delta z} p(L, z) \frac{dV}{d\Omega\,dz} \,dz \,d\Omega, 
\label{va1}
\end{eqnarray} where $p(L, z)$ is the function including all completeness mentioned before, $dV/(d\Omega\,dz)$ is the differential comoving volume at redshift z, and $\Delta\Omega$ represents the surveyed solid angle of all field regions covered by the M2FS pointings. The equation (\ref{va1}) can be expressed in a discrete form with respect to the solid angle $\Omega$, 
\begin{eqnarray}
V_a = \sum_j \Omega_j \int_{\Delta z} p_j(L, z) \frac{dV}{d\Omega\,dz} \,dz,
\label{va}
\end{eqnarray} where the sum is over all fields and $p_j(L, z)$ is the completeness function for the $j$th field. Table \ref{fieldsinfo} gives all values of $\Omega_j$ in Column 3. We plot $V_a$ as a function of \lya\ luminosity as the red line in the upper panel of Figure \ref{valf}. The blue line is for $z\approx5.7$ obtained in \pz. When a LAE is luminous enough, its $V_a$ reaches ${\sim}1.9\times10^6$ cMpc$^{-3}$. $V_a$ is then calculated for each LAE in our sample (Column 9 in Table \ref{glaes}), except the three bright ones in COSMOS1 and COSMOS3. Considering the survey limit, we set a faint-end cut at ${\rm log}_{10}L({\rm Ly\alpha})=42.6$ below which $V_a$ decreases significantly smaller than ${\sim}10^5$ cMpc$^3$. The remaining sample is divided by eight bins with equal sizes of 0.15 dex in a logarithmic $L$(\lya) range of $42.6{-}43.7$. The binned \lya\ LF for our sample is listed in Table \ref{binlf}. In the lower panel of Figure \ref{valf}, we plot $\phi({\rm log}L)$ in a logarithmic scale using the red symbols with error bars.

We parameterize the \lya\ LF at $z\approx6.6$ using a Schechter function \citep{schechter76}, 
\begin{eqnarray}
   \phi(L) = \frac{\phi^*}{L^*}\left(\frac{L}{L^*}\right)^{\alpha}{\rm exp}\left(-\frac{L}{L^*}\right),
   \label{lyalf0}
\end{eqnarray} where $\phi^*$ is the characteristic volume density, $L^*$ is the characteristic luminosity, and $\alpha$ is the faint-end slope. We fit the Schechter function with or without the three brightest bins (marked by w/o in Table \ref{lfpara}). The luminosity limit is not deep enough to constrain $\alpha$, so we fix $\alpha=-1.5$ for the consistency with \pz. This value is also the fiducial value from the literature \citep{mr04, kashikawa06, kashikawa11, ouchi08, ouchi10, hu10, zheng16}. The fitting results are listed in Table \ref{lfpara}.

\begin{deluxetable}{lcc}[t]
\tablecaption{The Schechter Parameters of our \lya\ LF at $z\approx6.6$
\label{lfpara}}
\tablehead{
\colhead{$\alpha$} & \colhead{log$_{10}\phi^*$} & \colhead{log$_{10}L^*$}\\
\colhead{} & \colhead{(cMpc$^{-3}$)} & \colhead{(erg~s$^{-1}$)}}
\startdata
\vspace{0.1cm}
 -1.5 (w) &  $-4.26_{-0.67}^{+0.29}$ &  $43.02_{-0.20}^{+0.40}$ \\\vspace{0.1cm}
 -1.5 (o) &  $-4.00_{-0.49}^{+0.29}$ &  $42.86_{-0.18}^{+0.22}$ \\
\enddata
\tablecomments{w/o refers to the fitting with or without the three brightest bins.}
\end{deluxetable}

\subsection{A Bright-end Bump}

\begin{figure}[t]
\epsscale{1.15}
\plotone{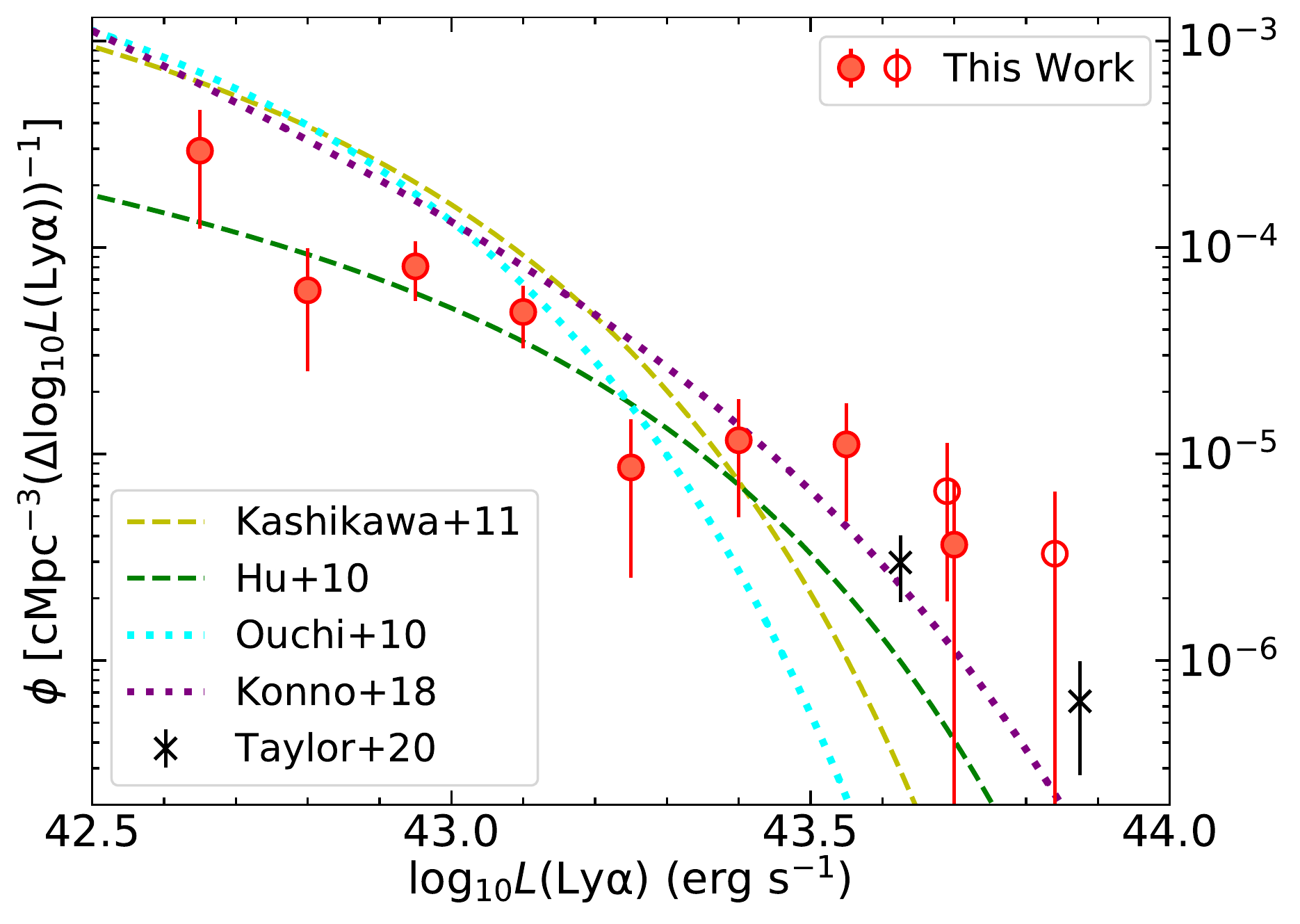}
\caption{\lya\ LFs at $z\approx6.6$. The red filled circles with error bars denote our binned LF data. The red lines are the best-fit Schechter function fixed at $\alpha=-1.5$. The red open circles indicate our LF including the two most luminous LAEs in our sample, which are horizontally shifted by 0.01 dex for clarity. The two crosses denote the \lya\ LF from \citet{taylor20}. The yellow and green dashed lines are the Schechter functions at a fixed $\alpha=-1.5$ from \citet{kashikawa11} and \citet{hu10}, respectively. The two studies used spectroscopically confirmed samples like us. The cyan and purple dotted lines are the Schechter functions with $\alpha=-1.5$ and $-2.5$ from \citet{ouchi10} and \citet{konno18}, respectively. The two studies used photometrically selected samples.
\label{lfs}}
\end{figure}

We compare our \lya\ LF at $z\approx6.6$ with previous studies in Figure \ref{lfs}. We plot four results of the \lya\ LF at $z\approx6.6$ from the literature. Two LFs are from spectroscopically confirmed samples \citep{hu10, kashikawa11}, while the other two are from photometrically selected samples \citep{ouchi10, konno18}. There exist large discrepancies between the measured LFs. The LFs based on photometric LAE candidates (the dotted lines in the Figure \ref{lfs}) agree with each other in the faint end, but deviate at the bright end. As for the studies using spectroscopically confirmed samples, the $z\approx6.6$ LF given by \citet{kashikawa11} is higher than that of \citet{hu10} by a factor of ${\sim}3$ at $10^{43.5}$ erg~s$^{-1}$.

Our \lya\ LF at $z\approx6.6$ agrees well with that of \citet{hu10}. Both of our work use spectroscopically confirmed LAEs over a total survey area of ${>}1$ deg$^2$. Most LAEs were found at $z=6.5$ using a different filter (NB912/9140\AA) in \citet{hu10}. The consistency between the results of ours and \citet{hu10} implies that the SDF field studied by \citet{kashikawa11} probably covers a dense region of ${\lesssim}L^*$ LAEs at the redshift. At the bright end, our \lya\ LF is higher because no LAE with $L({\rm Ly\alpha})>10^{43.5}$erg~s$^{-1}$ is found by \citet{hu10}. Our LF has an excess in the two brightest bins. We include the two LAEs in COSMOS1 (No.~29 and No.~36) to estimate a lower limit of the binned LF in the range of ${>}10^{43.5}$ erg~s$^{-1}$ which is defined as the ultraluminous (UL) range. We follow the same procedure in Section 4.2 to calculate their $V_a$ by enlarging the total survey area with an additional pointing COSMOS1. We show the estimated LF in the UL range indicated by the red open circles in the Figure \ref{lfs}. Such a deviation at the UL end implies that the Schechter function can not well describe the \lya\ LF at $z>6$.

At ${\rm log}_{10}L({\rm Ly\alpha}) \geq 43.5$, the bump feature exceeds the Schechter function at the bright end of the $z\approx6.6$ \lya\ LF. Such an excess of the \lya\ LF at the $z\approx6.6$ has been reported by previous studies \citep[e.g.,][]{ouchi10, matthee15, konno18}. \citet{ouchi10} has an excess LF data due to one exceptionally luminous LAE \citep[Himiko;][]{ouchi13}, which is the No.~28 LAE in our sample. \citet{matthee15} shows a clear excess at the bright end of the LF when including the bright LAEs in the wide SA22 field. \citet{konno18} discuss the systematic effects in the measurements to explain the bright-end bump. \citet{hu10} and \citet{kashikawa11} did not find the bright-end bump of the \lya\ LF at the $z\approx6.6$. \citet{hu10} used a large luminosity bin ($\sim$0.3 dex) that probably erases the bump feature. Based on a relatively small survey area, the cumulative \lya\ LF adopted by \citet{kashikawa11} does not show a bump feature.

The LF at the bright end is likely affected by cosmic variance due to a relative rarity of UL LAEs. In our sample, the seven LAEs in the three brightest LF bins locate in three different fields (SXDS, A370a, and COSMOS), which largely reduces the influence from the field-to-field variation. We also demonstrate the possibility of cosmic variance by comparing our result with that of \citet{taylor20}. Our spectroscopic survey find five UL LAEs in a total area of ${\sim}2$ deg$^2$ (a comoving volume of ${\sim}1.9\times10^6$ cMpc$^{-3}$). \citet{taylor20} has a total number of 11 in a much wider area of ${\sim}34$ deg$^2$ (a comoving volume of ${\sim}2.4\times10^7$ cMpc$^{-3}$) that can largely reduce the uncertainty from the cosmic variance. Their LF measurement, as shown by the black crosses in Figure \ref{lfs}, is higher than other results of previous studies. Our density measurement in the UL range is consistent with \citet{taylor20}. It reveals that the bright-end bump exists in the \lya\ LF at $z\approx6.6$. The LF bump reveals that the \lya\ LF may evolve differentially at the bright and faint end during the reionization. We give a further discussion in the next section. 

\section{Discussion}

\subsection{\lya\ LF Evolution and Ionized Bubbles}

\begin{figure}[t]
\epsscale{1.15}
\plotone{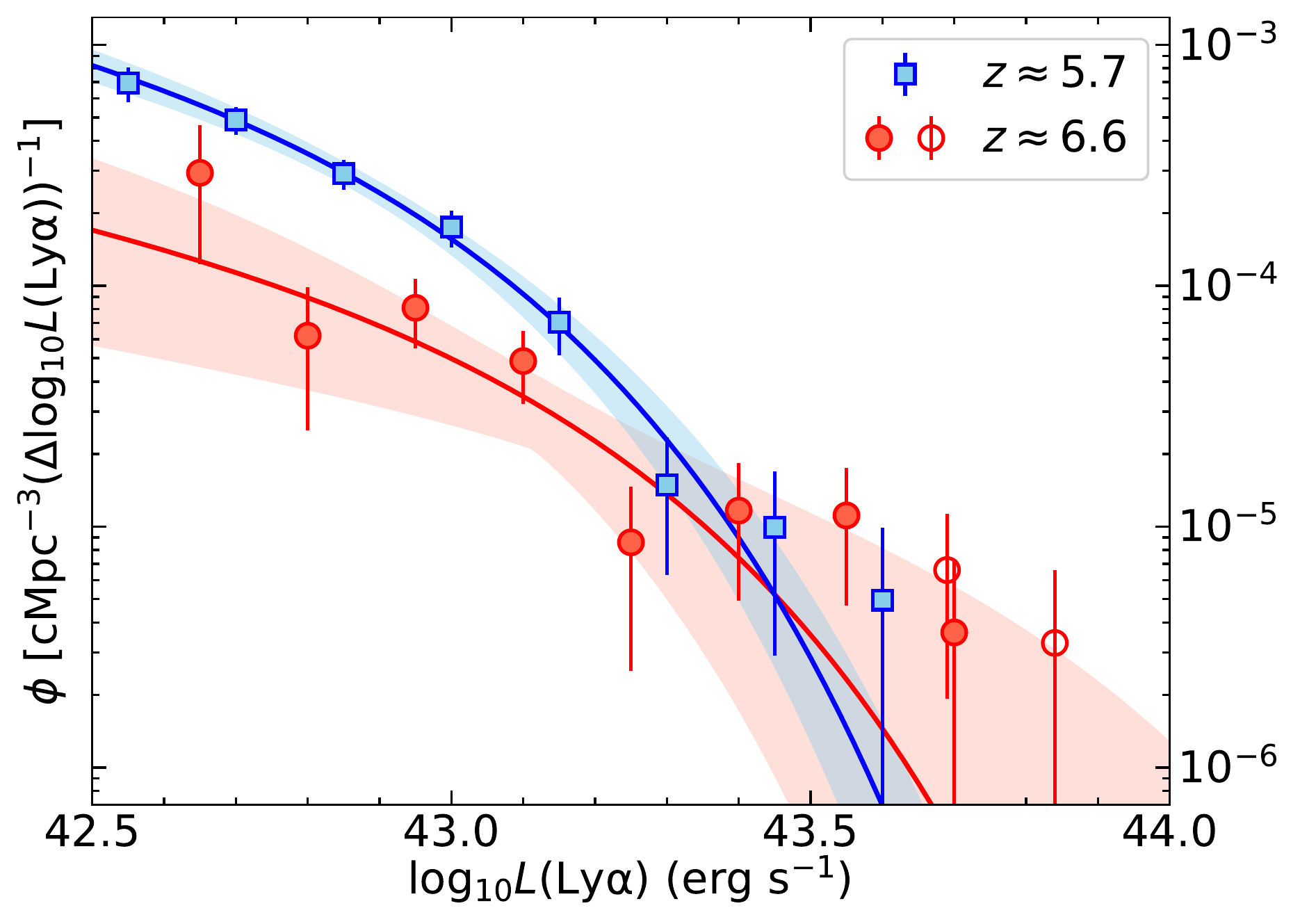}
\caption{The \lya\ LFs at $z\approx5.7$ (blue, from \pz) and $z\approx6.6$ (red, this work), which are both obtained by our spectroscopic survey. The lines represent the best-fit Schechter functions while the corresponding colored shades cover $1\sigma$ region of the fitting.
\label{evo}}
\end{figure}

We compare our $z\approx6.6$ \lya\ LF with the $z\approx5.7$ \lya\ LF from \pz\ in Figure \ref{evo}. The bright-end bump of the $z\approx6.6$ LF converges towards the $z\approx5.7$ LF. Meanwhile, a rapid LF evolution occurs at the faint end between the two redshifts. The \lya\ LF thus evolves diversely at the bright and faint ends. The \lya\ LF evolution at $z\gtrsim6$ is not only determined by galaxy evolution, but also affected by cosmic reionization. The bright-end bump probably links to the ionized bubbles, the \hii\ regions in the vicinity of galaxies during the EoR.

In the process of reionization, Lyman continuum photons emitted by young, massive stars escape from galaxies and ionize their surrounding IGM \citep{iliev06b}. The ionized bubbles form, grow and then coalesce. In such a patchy structure, brighter LAEs generally hold larger ionized bubbles \citep{yajima18}. \lya\ photons can redshift further away from the scattering resonance. The visibility of LAEs thus relates to their luminosity \citep{haiman05}. As the IGM becomes more neutral, the number density of observed faint LAEs descend remarkably while that of bright LAEs do not change notably \citep{matthee15, konno18, weinberger18, taylor21} leaving a bump feature at the bright end. At higher redshift, the difference between the faint- and bright-end LF evolution increases. The bright-end bump is indeed robustly detected in the \lya\ LF at $z\sim7.0$ by the LAGER survey \citep{zheng17, hu19}. They also show that the IGM has a higher \hf\ at $z\sim7.0$ than at $z\approx5.7$ and 6.6.

The large ionized bubbles of the luminous LAEs are supported by extraordinary strong UV radiation. One of the explanations is that these objects may hold active galactic nucleus (AGN) activity. Our $z\approx6.6$ \lya\ LF is constructed using the spectroscopically confirmed LAEs, excluding the possibility of strong AGN avctivities \citep[e.g.,][]{taylor20, taylor21, ning20}. The typical LAEs are star-forming galaxies with low metallicities and low dust. However, even the brightest LAE without an AGN fails to form a large ionized bubble at $z\approx6.6$ \citep{mr06, park21}. Therefore, these UL LAEs likely reside in significantly overdense regions in which a number of low-luminosity galaxies together generate enough ionizing photons to produce large H II bubbles.

The most luminous LAEs in our sample are complex systems with galaxy components. Some of them have been individually observed by the follow-up spectroscopy and/or deep IR imaging. For example, the two UL luminous LAEs, No.~28 and No.~29 (Himiko and CR7, respectively), are complex assembling systems of multiple components, which are revealed by the near-IR images from the {\it Hubble Space Telescope} ({\it HST}) \citep{ouchi13, sobral15}. No.~13 (MASOSA) has a compact \lya\ morphology but is undetectable in other available bands \citep{matthee19}. No evidence of AGN activity is found for all of them. Although CR7 and MASOSA (in COSMOS1) are only used to constrain a lower limit of the UL LF, they represent a typical LAE population at the bright end. We still need to conduct deep imaging observations to other bright LAEs, such as No.~33 in A370a, using the {\it HST} and/or the future {\it James Webb Space Telescope} ({\it JWST}). The {\it JWST} can also detect the potential high-ionization metal lines, such as C\,{\sc iv} $\lambda 1549$, He\,{\sc ii} $\lambda 1640$, and C\,{\sc iii}] $\lambda 1909$, of the high-redshift star-forming galaixes and help us to explore the nature of the very luminous LAEs and the ionized bubbles in the EoR.

\subsection{Neutral Hydrogen Fraction of the IGM at $z\approx6.6$}

By measuring the evolution of the \lya\ luminosity density $\rho_{\rm Ly\alpha}$, we constrain the neutral hydrogen fraction \hf. We use a method in the literature \citep[e.g.,][]{zheng17,konno18,hu19}. The density $\rho_{\rm Ly\alpha}$ is related to the UV luminosity density $\rho_{\rm UV}$ by the equation, 
\begin{equation}
  \rho_{\rm Ly\alpha} = \rho_{\rm UV}\, \kappa\, f_{\rm esc}\, T_{\rm IGM},
\end{equation} where $\rho_{\rm UV}$ is the intrinsic UV luminosity density from the stellar population within a galaxy, $\kappa$ is the conversion factor from UV to \lya\ luminosity that is determined by the stellar population, $f_{\rm esc}$ is the \lya\ escape fraction from a galaxy through the ISM \citep{dijkstra14}, and $T_{\rm IGM}$ is the IGM transmission of \lya\ emission. Under the assumption that the physical properties are similar between the LAEs at $z\approx5.7$ and $z\approx6.6$, we can obtain the \lya\ transmission ratio 
\begin{equation}
  \frac{T_{\rm IGM}^{z\approx6.6}}{T_{\rm IGM}^{z\approx5.7}}=\frac{\rho_{\rm Ly\alpha}^{z\approx6.6}/\rho_{\rm Ly\alpha}^{z\approx5.7}}{\rho_{\rm UV}^{z\approx6.6}/\rho_{\rm UV}^{z\approx5.7}}.
\end{equation} \citet{bouwens21} has improved determinations of the UV LF. We use the Schechter parameters in their Table 5 to calculate $\rho_{\rm UV}$ at each redshift and obtain $\rho_{\rm UV}^{z\approx6.6}/\rho_{\rm UV}^{z\approx5.7}=0.65\pm0.06$ by interpolation.

\begin{figure}[t]
\epsscale{1.15}
\plotone{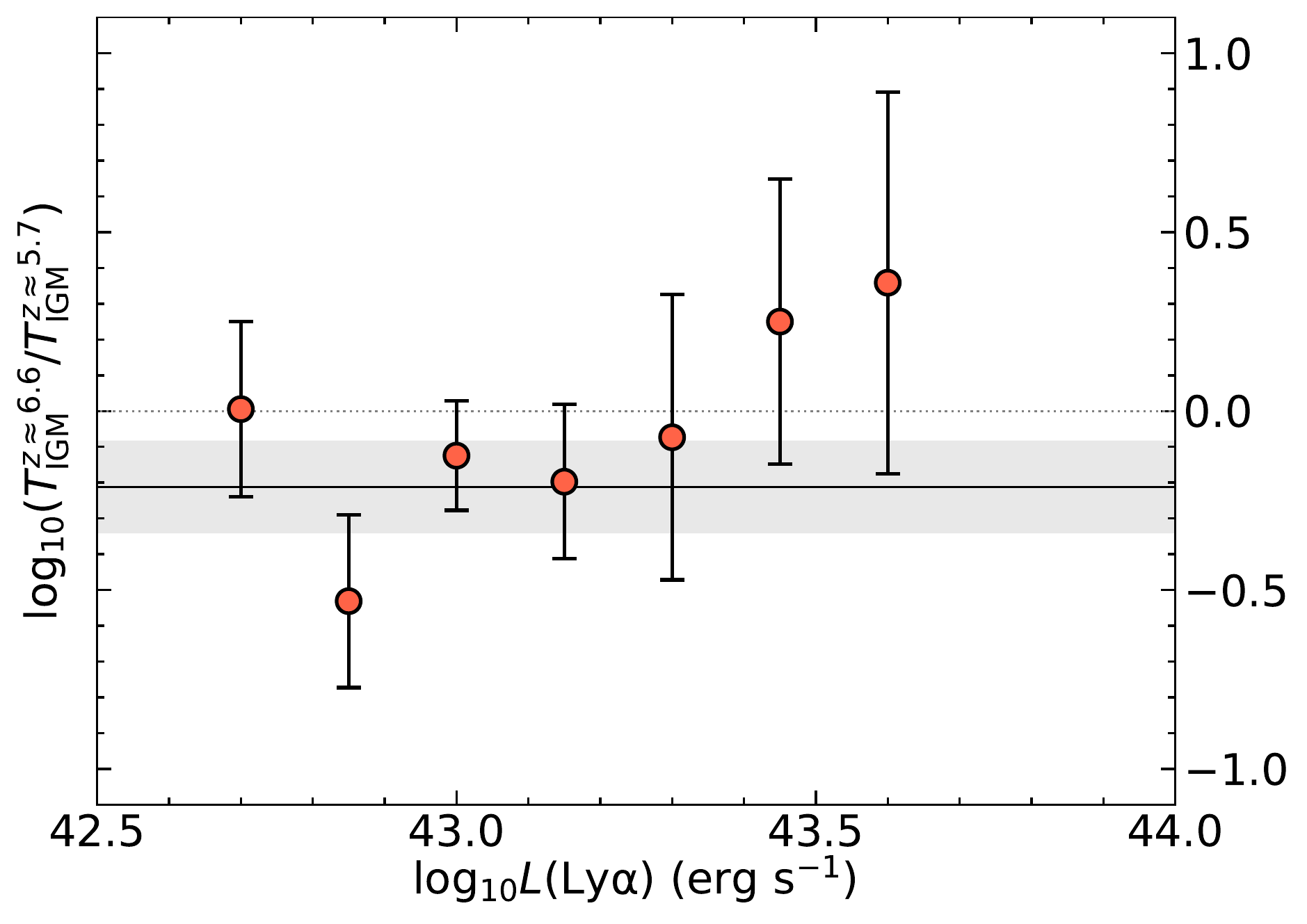}
\caption{${T_{\rm IGM}^{z\approx6.6}}/{T_{\rm IGM}^{z\approx5.7}}$ as a function of ${\rm log}_{10}L({\rm Ly\alpha})$. The horizontal dotted line represents the unchanged \lya\ IGM transmission factor from $z\approx5.7$ to $z\approx6.6$. The shaded region denotes an overall estimation of ${T_{\rm IGM}^{z\approx6.6}}/{T_{\rm IGM}^{z\approx5.7}}=0.61\pm0.18$.
\label{igmtran}}
\end{figure}

We calculate the \lya\ $T_{\rm IGM}$ ratio for each log$_{10}L$(\lya) bin. Assuming $\rho_{\rm UV}^{z\approx6.6}/\rho_{\rm UV}^{z\approx5.7}$ is constant, we obtain ${T_{\rm IGM}^{z\approx6.6}}/{T_{\rm IGM}^{z\approx5.7}}$ as a function of ${\rm log}_{10}L({\rm Ly\alpha})$. The result is shown in Figure \ref{igmtran}. The $T_{\rm IGM}$ ratio is higher (${\sim}1$) at brighter range (log$_{10}L$(\lya) $\gtrsim43.4$) than fainter. The transmission factor $T_{\rm IGM}^{z\approx5.7}$ corresponds to a highly ionized IGM with \hf\ $\lesssim10^{-4}$ in the completed reionization stage at $z\lesssim6$ \citep{fan06}, which can be treated as a normalization here. This figure thus demonstrates that \lya\ photons emitted from bright LAEs encounter more, or even fully ionized IGM than faint LAEs. The trend naturally reveals the scenario of the ionized bubbles whose size becomes larger as the \lya\ transmission increases \citep{mr06, dijkstra07b}. For a further \hf\ estimation, we also measure an averaged $T_{\rm IGM}$ ratio. To measure the overall $\rho_{\rm Ly\alpha}$ for the two redshifts, we use cubic spline functions to interpolate the binned \lya\ LFs and integrate them over the ${\rm log}_{10}L({\rm Ly\alpha})$ range overlapped by the two data sets (${\sim}42.6{-}43.6$). The \lya\ luminosity density ratio is $\rho_{\rm Ly\alpha}^{z\approx6.6}/\rho_{\rm Ly\alpha}^{z\approx5.7}=0.40\pm0.11$. We obtain ${T_{\rm IGM}^{z\approx6.6}}/{T_{\rm IGM}^{z\approx5.7}}=0.61\pm0.18$ which is denoted by the shaded region in Figure \ref{igmtran}.

We use the IGM transmission ratio to estimate \hf\ by comparing with theoretical models. We first interpolate the analytic model of \citet[][Figure 25]{santos04} by adopting a \lya\ velocity shift of ${\sim}200$ km~s$^{-1}$ as a lower limit (at $z\sim2$; \citealt{matthee21}). The ratio ${T_{\rm IGM}^{z\approx6.6}}/{T_{\rm IGM}^{z\approx5.7}}$ is then converted to \hf\ $\sim0.2-0.6$. This ratio also corresponds to an \hii\ bubble size of ${\sim}1{-}10$ pMpc \citep{dijkstra07b}. The range of the bubble size yields \hf\ $\sim0.1-0.4$ by comparing it as a function of the mean ionized fraction \citep{furlanetto06}. We give a conservative estimation of \hf\ $\sim0.3\pm0.1$. This result reveals that the IGM environment is still fairly neutral at $z\approx6.6$ and the reionization process may be rapid and rather late \citep[e.g.,][]{weinberger18, becker21, davies21}.

\section{Summary}

We have presented a new sample of 36 LAEs at $z\approx6.6$ from our Magellan M2FS spectroscopic survey of high-redshift galaxies. The candidates were selected from the narrowband NB921 photometry and broadband photometry. The whole sample was covered by 13 M2FS pointings with a total sky area of $\sim$2 deg$^2$. The on-source integration time was $\gtrsim5$ hrs per pointing. The LAEs were identified based on the 1D and 2D M2FS spectra. We measured LAE redshifts by fitting a composite \lya\ line template to the individual 1D lines. The secure redshifts were used with the NB921 and $z'$ band photometric data to derive their \lya\ luminosities. These LAEs span a \lya\ luminosity range of $\sim 3\times10^{42} - 7\times10^{43}$ erg~s$^{-1}$, including some of the most luminous LAEs known at $z\gtrsim6$. We revealed a positive correlation between the line width and luminosity of \lya\ emission at $z\approx6.6$ like that at $z\approx5.7$.

Using our spectroscopic LAE sample, we have obtained the \lya\ LF at $z\approx6.6$, after we considered a comprehensive sample completeness correction. A clear bump was found at the bright end of the \lya\ LF, which is probably caused by ionized bubble structures around very luminous LAEs that reside in overdense regions. We compared the \lya\ LF at $z\approx6.6$ with the $z\approx5.7$ LF from \pz, and confirmed a rapid evolution at the faint-end LF. But there is a lack of evolution at the bright end. From such an evolution, the measured fraction of neutral hydrogen in the IGM at $z\approx6.6$ is roughly \hf\ $\sim0.3\pm0.1$. We emphasize that the above results are based on the spectroscopically confirmed LAE samples. Our Magellan M2FS survey has produced a spectroscopic sample of $\sim$300 high-redshift galaxies  (36 LAEs at $z\approx6.6$ in this work and 260 LAEs at $z\approx5.7$ from \pn). This sample provides unique targets for further studies of the high-redshift universe.

\acknowledgments
We acknowledge support from the National Key R\&D Program of China (2016YFA0400703), the National Science Foundation of China (11721303, 11890693), the China Manned Space Project (CMS-CSST-2021-A05, CMS-CSST-2021-A07), and the Chinese Academy of Sciences (CAS) through a China-Chile Joint Research Fund \#1503 administered by the CAS South America Center for Astronomy. Z.Y.Z. acknowledges support by the National Science Foundation of China (11773051, 12022303), the China-Chile Joint Research Fund and the CAS Pioneer Hundred Talents Program. This paper includes data gathered with the 6.5 meter Magellan Telescopes located at Las Campanas Observatory, Chile.

\facilities{Magellan: Clay (M2FS).}
\bibliography{ms}
\end{document}